**Robust Electrocaloric Performance Enabled by Highly-Polar Frustrated Nanodomains in $NaNbO_3$-Based Ferrodistortive Relaxor**

Feng Li, Changshun Dai, He Qi*, Jiecheng Liu, Xiaoming Shi*, Heng Zhou, Qiong Yang*, Mingsheng Long, Lei Shan, Chunchang Wang, Jianli Wang, Zhenxiang Cheng*

F. Li, C.S. Dai, M.S. Long, L. Shan

Institutes of Physical Science and Information Technology, Leibniz International Joint Research Center of Materials Sciences of Anhui Province, Anhui University, Hefei 230601, China.

H. Qi

School of Materials Science and Engineering, Hainan University, Haikou, Hainan, 570228, China.

Email: qihe@hainanu.edu.cn

J.C. Liu, X.M. Shi

Department of Physics, University of Science and Technology Beijing, Beijing, 100083, China.

Email: xmshi@ustb.edu.cn

H. Zhou, Q. Yang

Hunan Provincial Key Laboratory of Thin Film Materials and Devices, School of Materials Science and Engineering, Xiangtan University, Xiangtan, Hunan 411105, China.

Email: qyang@xtu.edu.cn

C.C. Wang

Laboratory of Dielectric Functional Materials, School of Materials Science and Engineering, Anhui University, Hefei, 230601, China.

J.L. Wang, Z.X. Cheng

Institute for Superconducting and Electronic Materials, Australian Institute for Innovative Materials, University of Wollongong, North Wollongong 2500, Australia.

Email: cheng@uow.edu.au

**Abstract**

Solid-state refrigeration technologies, represented by electrocaloric effect (ECE), are renowned for zero global-warming-potential and high cooling efficiency. Synergistically achieving high electrocaloric effect ($\Delta T$) and wide temperature span ($\Delta T_{span}$) for EC materials takes a leapfrog toward practical cooling applications, typical for integrated circuits. Guided by phase-field simulation, Ba(Ti, Hf)$O_3$ dubbed as a polar wrench, establishes polar frustration by setting up local stress field and manipulating octahedral oxygen tilt (OOT) in $NaNbO_3$-based relaxor. The resultant *P4bm* framework entails short-range and highly-polar ferrodistortive nanodomains, i.e., the abundant highly-polar nanodomains facilitate to increase entropy change and robust OOT enables to impede thermal perturbations. Consequently, a large $\Delta T$ of 0.85 K and 0.70 K with an ultrawide $\Delta T_{span}$ of 118 K and 130 K is obtained, contributing to an ultrahigh figure of merit of > 90 $K^2$ in $NaNbO_3$-Ba(Ti, Hf)$O_3$, significantly outperforms its counterparts. The local structure responsible for robust EC performances are decrypted through 2D information from atomic-resolution scanning transmission electron microscope, 3D big-box model constructed from neutron total scattering and DFT calculations. These findings highlight that polar frustration strategy in ferrodistortive relaxor enables to pioneer emergent EC performances, and also unearth potential entropy-change-based ferroelectric and ferromagnetic materials beyond.

**Keywords:** Electrocaloric effect; $NaNbO_3$-based; Oxygen octahedral tilt; Polar frustration; Ferrodistortive relaxor.

## 1. Introduction

Solid-state refrigeration technologies, represented by electrocaloric effect (ECE), garnered significant attentions for their high efficiency, compact integration, and zero global-warming-potential.[1] Compared to the widely-used vapor-compression technologies, ECE offers distinct advantages for localized and manageable cooling in advanced microchips and wearable devices. EC researches experienced decades of hiatus after a giant ECE was discovered in inorganic film and polymer with a large temperature change ($\Delta T$) of 12 K around (anti)ferroelectric-paraelectric transition.[2] Since ECE is proportional to entropy change ($\Delta S$) and it is facilitated by dipole change for disorder-to-order states associated with structural transitions, field-induced antiferroelectric/relaxor-to-ferroelectric transformation boosts a large $\Delta T$, *e.g.*, in $PbZrO_3$-/$Pb(Mg_{0.5}W_{0.5})O_3$,[3] $Bi_{0.5}Na_{0.5}TiO_3$-,[4] $K_{0.5}Na_{0.5}NbO_3$-,[5] and $BaTiO_3$-based ceramics.[6] These advances underscore the potential of EC materials for next-generation cooling applications.

Practical implementation, especially in integrated circuit cooling, requires not only a large $\Delta T$ but also a wide temperature span ($\Delta T_{span}$, defined as the range where $\Delta T \geq 80\%$ of $\Delta T_{max}$) to ensure sufficient refrigeration capacity.[7] While first-order ferroelectrics like $BaTiO_3$ exhibit large $\Delta T$ due to high latent heat, they inherently suffer from a narrow $\Delta T_{span}$.[8] Strategies such as laminated composites or multiphase coexistence regions–including invariant critical points (ICP) have been explored to extend the operational temperature window near room temperature.[9] Although these approaches improve temperature stability, they often do so at the expense of reduced $\Delta T$. Alternatively, for typical $Bi_{0.5}Na_{0.5}TiO_3$-based relaxors, nonergodic-to-ergodic phase regulation can yield a thermally-stable EC profile. However, aliovalent doping often introduces cation/oxygen vacancies that disrupt long-range ferroelectric domains, the side effect of inhibition for domain-wall motions is nevertheless accompanied. This results in a sharply suppressed polarization and diminished $\Delta T$.[10] Moreover, the delicate balance between $\Delta T$ and $\Delta T_{span}$ in relaxors is easily disturbed by thermal energy: polar nanoregions (PNRs) become misaligned above the freezing temperature ($T_f$) or frozen below it, either of which degrades the EC response.[11] Thus, conventional approaches relying on structural heterogeneity face fundamental challenges.

According to thermodynamics and statistics, entropy change $\Delta S$ is expressed as:

$$\Delta S = \frac{\ln \Omega}{\varepsilon_0\, \Theta} \Delta P^2 \qquad \text{(Eq. 1)}$$

where $\Theta$ reflects the dipole correlation strength, $\Omega$ is the number of polarization directions of polar state, and $\Delta P$ denotes the change in polarization response.[12] Maximizing $\Delta S$–hence $\Delta T$, requires a large $\Omega$, a high $\Delta P$, and a low $\Theta$. This underscores the criticality in making matrix ferroelectricity (high $\Delta P$) intact in the process of establishing hashed polar nanodomains (PNDs, large $\Omega$ and low $\Theta$). To ensure thermal stability of PNDs, the energy barrier for polarization flipping between adjacent states must be sufficiently high. This is hardly achieved in conventional relaxors since polar nanoregions with pseudocubic symmetry is highly vulnerable to thermal shock.[13] Recently, coupling of oxygen octahedral tilt (OOT) to electric properties has emerged as a promising route to novel functional behaviors.[14] In $NaNbO_3$ (NN)-based piezoelectrics, for example, buckling inter-octahedral B-O-B bond enhances piezoelectric performance and thermal stability, with OOT conferring greater resistance to thermal perturbation compared to morphotropic/polymorphic phase boundaries.[15] OOT has also been shown to modulate dielectric relaxation and ferroelectric behaviors.[16] Inspired by these findings, OOT is selected as tunable degree of freedom for optimizing EC performance. Constructing PNDs with an OOT-stabilized framework could simultaneously enhance entropy change (via PNDs alignment) and sustain ferroelectricity under thermal agitation (via the OOT framework). Such a structure is expected to enable concurrent improvement in both $\Delta T$ and $\Delta T_{span}$.

Guided by phase-field simulation, we propose a polar frustration strategy for the first time by incorporating strong polar cations, dubbed as a polar wrench. Through implementing this, the OOT are increasingly wrenched in tandem with frustration of long-range orders to short-range one, and an OOT framework with PNDs is thus built. Ultimately, an excellent EC performance with a large $\Delta T$ of 0.70/0.85 K and an ultrawide $\Delta T_{span}$ of 130/118 K with a record-high figure of merit (FOM) > 90 $K^2$ is obtained in $NaNbO_3$-$Ba(Ti, Hf)O_3$ system by elaborately controlling OOT and polar configuration. This work unravels a *terra incognita* of utilizing highly polar frustrated nanodomains for boosting EC performances in ferrodistortive relaxor.

## 2. Results and Discussion

### Design strategy and phase-field simulation for polar frustration

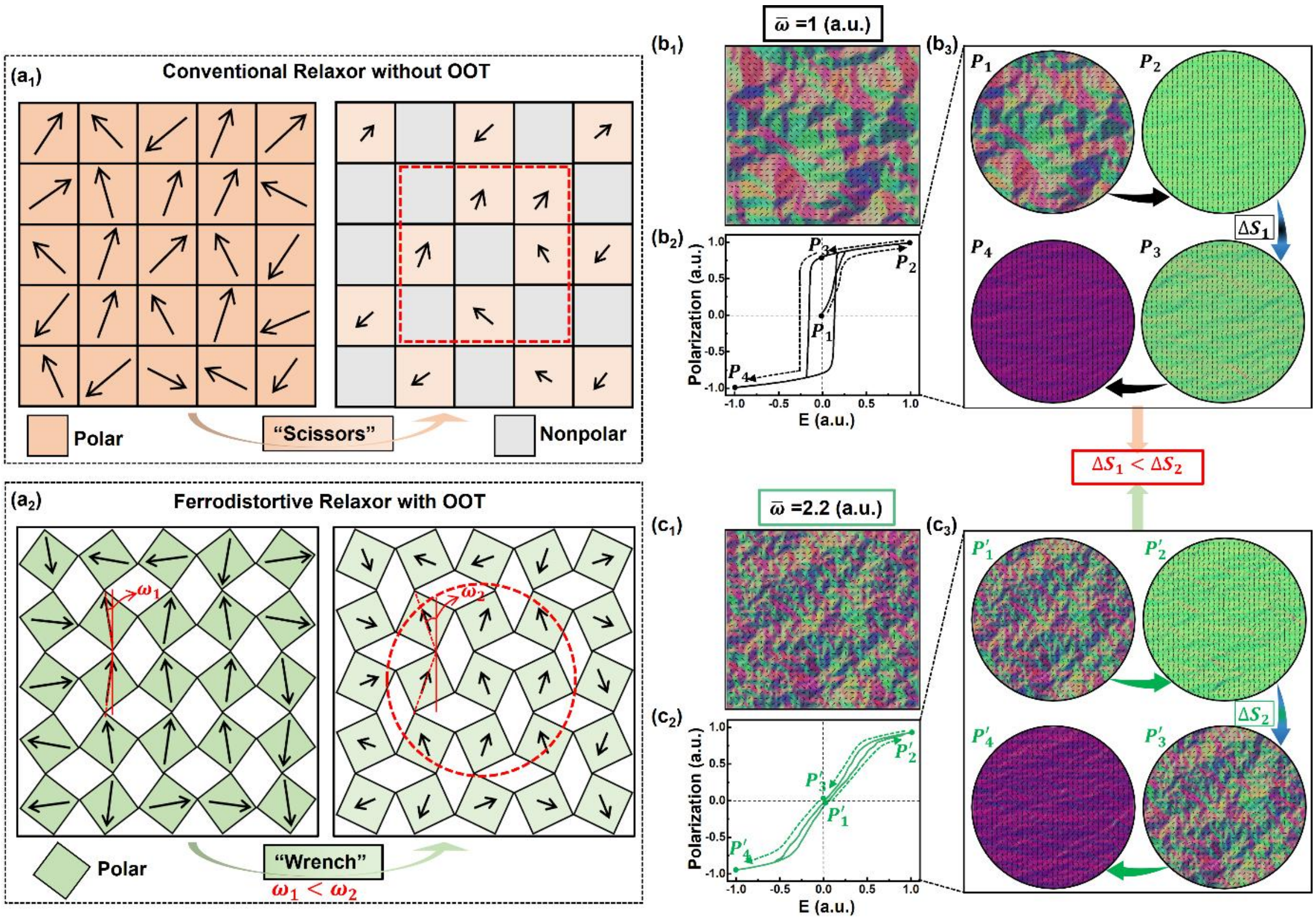

**Figure 1.** a1) Long-range order transforms to PNRs (indicated by square) in conventional relaxor without OOT. Dopant acts as scissors-like role since non-polar regions appear; a2) Long-range order transforms to PNDs (indicated by circle) in ferrodistortive relaxor with OOT. Polar frustrations occur via increasing OOT degree by strong polar cations (act as polar wrench role). The role of wrench instead of scissors is named here since PNDs are fully filled within the matrix. OOT degree is exaggerated here for clarity. b1, c1) Simulation of the 2D vector contours and phase distributions; b2, b3, c2, c3) Comparative display of simulated *P-E* loops and local structure evolution for $\bar{\omega}$ = 1.0 and 2.2 states during electric-field loading and unloading process (as indicated by arrows). The simulation graphs for $\bar{\omega}$ = 1.4 and 1.8 are displayed in Figure S1a, b in Supporting Information; Clearly, entropy change for $\bar{\omega}$ = 2.2 sample ($\Delta S_2$) is significantly larger than that of $\bar{\omega}$ = 1.0 ($\Delta S_1$). The phase field simulation details are described in Note S1 in Supporting Information.

Instead of the dopant acting as scissors-like role in partitioning long-range order to polar nanoregions (PNRs) within nonpolar matrix (conventional relaxor without OOT, Figure 1a1), the strong polar cations as a polar wrench frustrates long-range order into PNDs but with high ferroelectricity during ferroelectric-to-relaxor crossover (ferrodistortive relaxor with OOT, Figure 1a2). Clearly, establishing this polar frustration strategy will meet the roadmap target for high EC performance (Eq. 1). Phase-field simulation is thus performed to predict the potential of frustrating polar configuration in NN-based system. To clearly observe the polar configuration change, we choose NN ferroelectrics as an initial state (irreversible transformation of AFE-to-FE state for pure NN) and the standard deviation of normal distribution for OOT angle $\bar{\omega}$ is chosen as a variable. As displayed in Figure 1b$_1$, b$_2$, long-range ferroelectric order with a square *P-E* loop is observed. With a gradual increase in $\bar{\omega}$ and stress field, domains size remarkably decreases and long-range polar order frustrates, resulting in a gradual crossover of ferroelectric-to-relaxor with a near-zero remnant polarization $P_r$ (Figure 1c$_1$, c$_2$). These polar configurations strongly impact on entropy change, and their evolutions with loading and unloading *E* are presented, as selecting $\bar{\omega}$ = 1.0 and 2.2 samples as an example (Figure 1b$_3$, c$_3$). Upon applying *E*, the initial disorder state is triggered into an order state for $\bar{\omega}$ = 1.0 and it deviates a little degree with releasing *E* ($P_1 \rightarrow P_3$ in Figure 1b$_3$), resulting in a quite low entropy change $\Delta S_1$. Instead, the frustrated polar nanodomains induced by large $\bar{\omega}$ and stress field are responsive to *E*, and the mutual transition between long-range order and polar nanodomains is highly reversible during *E* loading and unloading process ($P_1' \rightarrow P_3'$ in Figure 1c$_3$). Encouragingly, compared to conventional polarization heterogeneity strategy, the maximal polarization intensity ($P_m$) remains unchanged during this process by designing polar configuration frustration (Figure S1c). Since $\Delta S$ is proportional to $P_m^2 - P_r^2$, a large $P_m$ and near-zero $P_r$ is beneficial to obtain high entropy change $\Delta S_2$. Consequently, frustrating polar configuration through OOT regulation possess a potent potential to enhance EC performance in NN-based materials.

Guided by the phase-field simulation, compared to the host $Na^+$ (1.39 Å) and $Nb^{5+}$ (0.64 Å) at *A*- and *B*-site, a larger size of $Ba^{2+}$ (1.61 Å) and smaller size of $Ti^{4+}$ (0.605 Å) is alloyed. $Ba^{2+}$ will impose a compressive stress and $Ti^{4+}$ a tensile stress (Figure S2), acting as a polar wrench. $Ba(Ti_{0.89}Hf_{0.11})O_3$ composition is designed as it resides at invariant critical point with ultrahigh dielectric activity, benefitting to elevating EC response.[9b, 17] Crucially, polar-wrench $Ba^{2+}$ and $Ti^{4+}$ cations with high-polarizability (6.4/2.93 Å$^3$) will not arouse strong frozen-in random electric field but primarily perturb the correlation of –(Na)Nb–(Na)O–(Na)Nb–O– chains (as indicated by dash parallelogram in Figure S2)[18], thereby maintaining the intrinsic polarity of the matrix. These strategies readily lead to a polar frustration with varied distribution of OOT.

## High EC performance for NN-BTH system

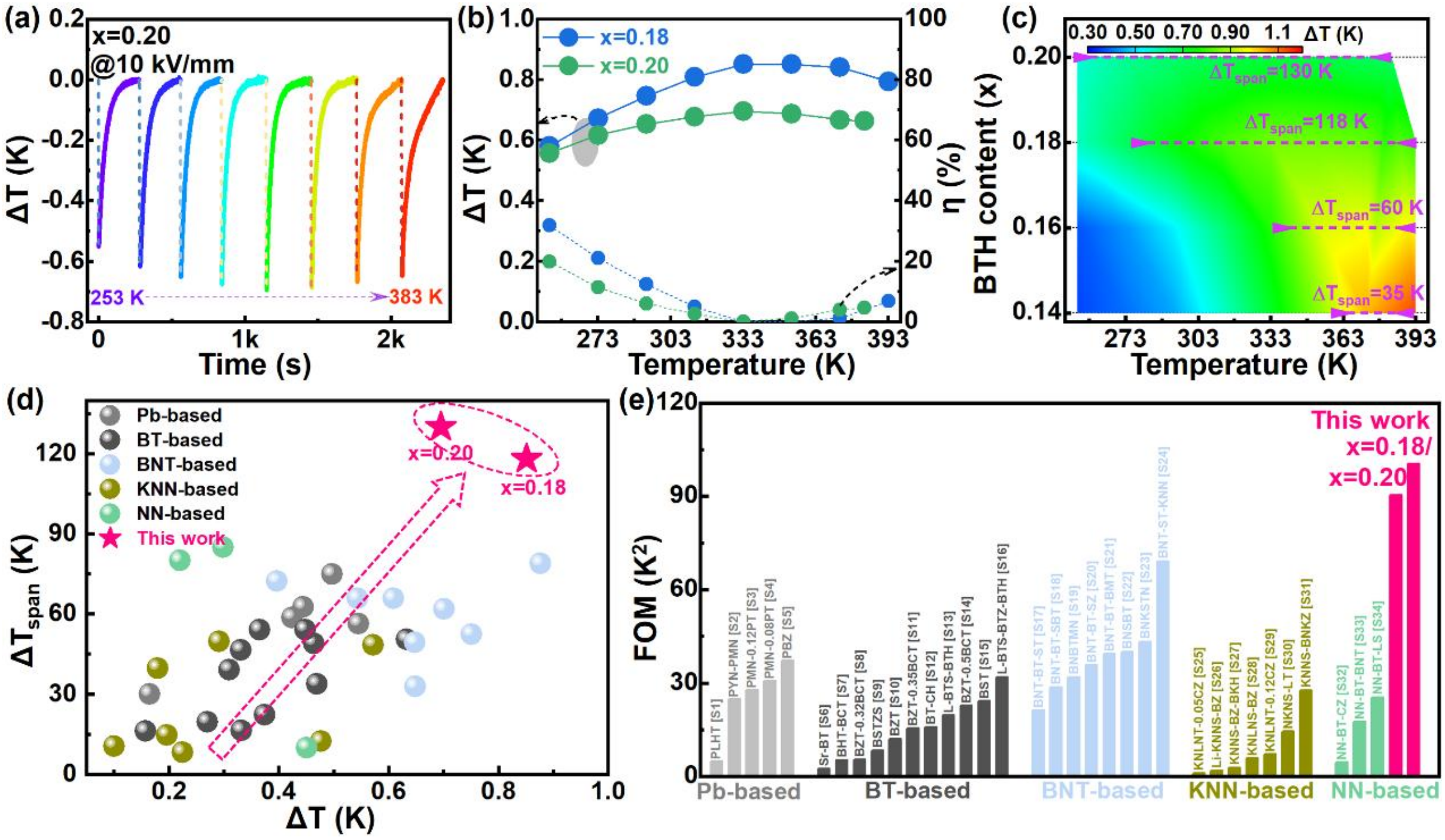

**Figure 2.** a) Directly measured ECE with endothermic peaks over 253 – 383 K at $E$ = 10 kV mm$^{-1}$ for $x$ = 0.20, the EC profiles for $x$ = 0.14 – 0.18 are displayed in Figure S3. b) Temperature dependence of $\Delta T$ and instability $\eta$ for $x$ = 0.18 and 0.20 samples. c) Contour map of $\Delta T$ and the temperature span for $\eta \leq 20\%$ are indicated; comparison of EC performance of d) $\Delta T$–$\Delta T_{span}$ and figure of merit (FOM) in this work with previously reported lead-based/free counterparts. Ref. [S1-S34] are listed in Supporting Information.

The introduced polar frustrations directly translate into remarkable EC performance. Temperature-dependent EC profiles for $x$ = 0.14 – 0.20 compositions are shown in Figure 2a and Figure S3. All samples exhibit characteristic endothermic peak upon $E$ is released, confirming a normal ECE. Notably, compositions with $x$ = 0.18 and 0.20 achieve peak $\Delta T$ values of 0.85 K and 0.70 K, respectively, under an $E$ of 10 kV mm$^{-1}$, with a dome-shaped temperature dependence that signifies excellent thermal stability (Figure 2b). This leads to an ultrawide operating temperature span ($\Delta T_{span}$) of 118 K for $x$ = 0.18 and 130 K for $x$ = 0.20, defined at $\eta \leq 20\%$ (where $\eta = \frac{\Delta T_{max} - \Delta T}{\Delta T_{max}} \times 100\%$), as shown in Figure 2c. The combination of large $\Delta T$ and unprecedented $\Delta T_{span}$ significantly surpasses the performance of other lead-free and lead-based ceramic systems (Figure 2d). Such concurrent enhancement in both key parameters is essential for practical applications, particularly in integrated circuit cooling. A relevant metric is the figure of merit (FOM), defined as $\Delta T \times \Delta T_{span}$, which reflects the theoretical refrigeration capacity of EC materials.[19] As compiled in Figure 2e, most reported

systems–including laminated $BaTiO_3$-based and $Bi_{0.5}Na_{0.5}TiO_3$-based ceramics, exhibit FOM values below 45 $K^2$. [9c, 20] In contrast, our $x$ = 0.18 and 0.20 samples achieve a record-high FOM of ≥ 90 $K^2$, representing a notable advance toward viable solid-state cooling technologies. These results unambiguously demonstrate that the long-standing challenge of synergistically improving $\Delta T$ and $\Delta T_{span}$ has been successfully addressed in the NN-BTH system. The superior EC performances validate the effectiveness of the polar frustration strategy, motivating further investigation into the underlying local structural mechanism.

**Domain structure and dielectric property**

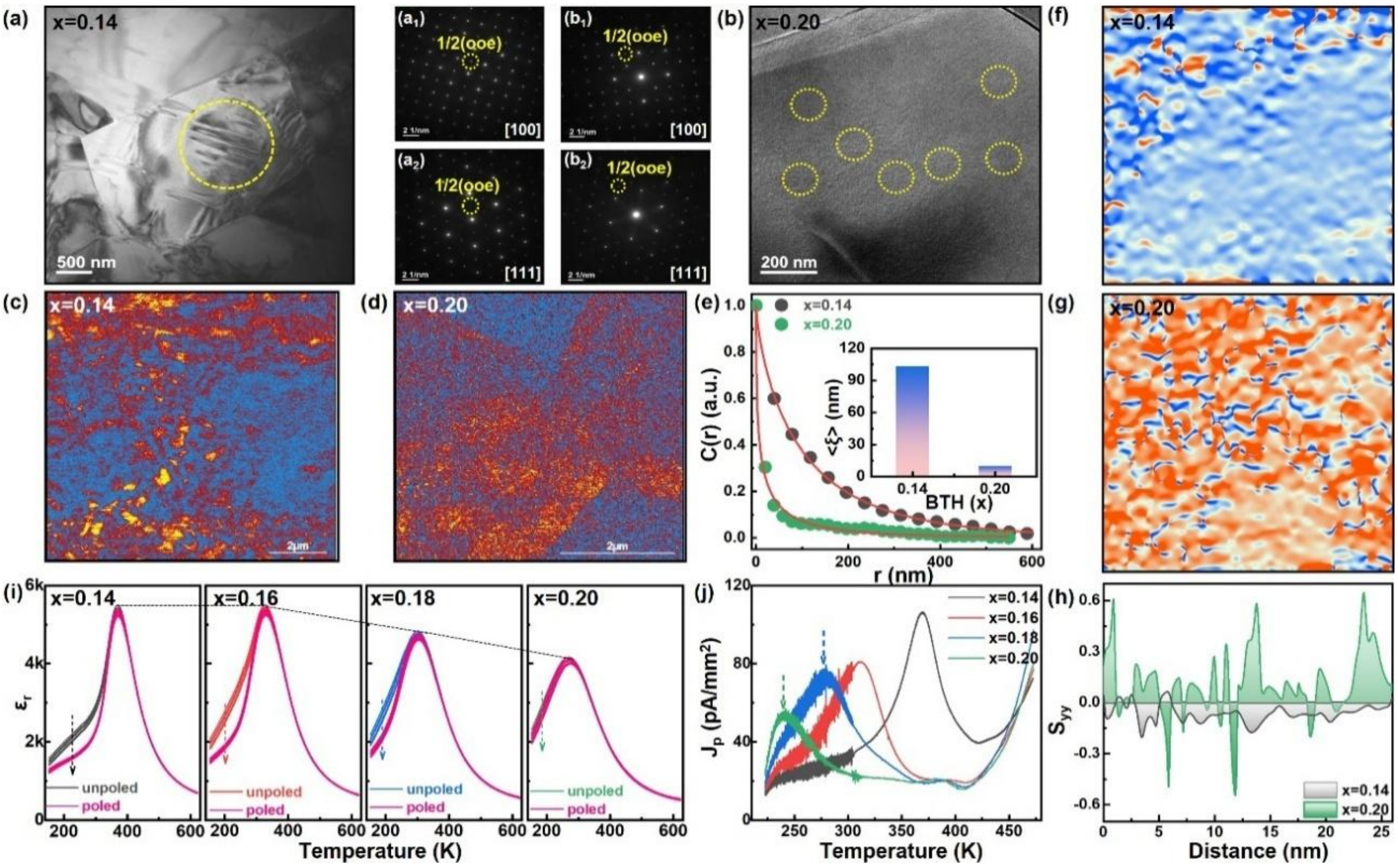

**Figure 3.** a, b) Bright-field transmission electron microscopy (TEM) images and $a_1$, $a_2$, $b_1$, $b_2$) selected-area electron diffraction (SAED) patterns along $[100]_c$ and $[111]_c$ directions for $x$ = 0.14 and 0.20 ceramics. TEM and SAED patterns for $x$ = 0.16 and 0.18 are displayed in Figure S4. c, d) Piezoelectric force microscopy (PFM) amplitude images for $x$ = 0.14 and 0.20 ceramics; piezoresponse amplitude profiles and local poling procedure are shown in Figure S5. e) Average autocorrelation function *C(r)* and corresponding fitting curves for $x$ = 0.14 and 0.20 compositions. The comparison of average short-range correlation length <$\xi$> are displayed in its inset. The $C(r)$ is defined as $\sigma^2 \times \exp\,[-(r/<\xi>)^{2b}]$, where parameter $r$ is distance from the central peak in $C(r_1, r_2)$ map, $b$ represents the roughness of polarization interface and <$\xi$> denotes the average short-range correlation length. f, g) Geometric phase analysis (GPA) of stress map ($S_{yy}$) for $x$ = 0.14 and 0.20 compositions and h) stress value profiles extracted from (f, g). i) Temperature dependence of $\varepsilon_r$ for unpoled and poled $x$ =0.14 − 0.20 samples within frequency of 1 − 100 kHz with heating cycles. j) Thermally stimulated depolarization current (TSDC) spectra for $x$ = 0.14 – 0.20.

To corroborate the domain structure evolution upon BTH incorporation, bright-field transmission electron microscopy (TEM) images and selected-area electron diffraction (SAED) patterns are obtained along the $[100]_c$ and $[111]_c$ zone axes for $x = 0.14 - 0.20$. Well-defined striped domains are evident in $x = 0.14$, which progressively break into blotchy, short-range PNDs in $x = 0.20$, as highlighted by yellow circles in Figure 3a, b. Notably, this nanoscale domain fragmentation does not alter the overall structural symmetry. SAED patterns consistently reveal (*ooe*)/2-type superlattice reflections along both $[100]_c$ and $[111]_c$ directions (Figure 3a$_1$-b$_2$, Figure S4), confirming the preservation of the *P4bm* phase with an $a^0a^0c^+$ OOT across all compositions. Piezoelectric force microscopy (PFM) further corroborates this morphological transition: large, island-like domains in $x = 0.14$ are replaced by a slush-like arrangement of PNDs in $x = 0.20$ (Figure 3c, d). The corresponding piezoresponse amplitude changes from a regular, ordered profile to a chaotic pattern (Figure S5a, b), underscoring the distribution of long-range correlation. This domain refinement profoundly impacts polar correlations and mobility. Analysis of the autocorrelation functions *C(r)* yields the short-range correlation length $<\xi>$,[21] which drops dramatically from 103 nm in $x = 0.14$ to just 10 nm in $x = 0.20$ (Figure 3e). This reduced correlation length significantly enhances domains reversibility, as visualized in local poling experiments (Figure S5c, d). After applying 10 and 15 V to a 3 × 3 μm$^2$ area, $x = 0.14$ retains a clear PFM contrast, indicating stable ferroelectric switching and correlating with a large remnant polarization $P_r$ and low entropy change between states $P_2$ and $P_3$ (Figure 1b$_3$). In contrast, $x = 0.20$ shows a faint, poorly retained contrast, revealing that the weakly correlated PNDs rapidly relax once poling field is removed–consistent with a high entropy change between $P_2'$ and $P_3'$ states (Figure 1c$_3$). Additionally, geometric phase analysis (GPA) directly visualizes the local stress field introduced by the polar-wrench effect (Figure 3f, g). The stress profile ($S_{yy}$) for $x = 0.20$ exhibit pronounced non-uniformity and higher maximum values compared to $x = 0.14$ (Figure 3h), confirming that BTH incorporation generates substantial random stress, which in turn drives polar order fragmentation. This aligns with reported roles of elastic stress in promoting short-range domain ordering to enhance functional properties.[22]

The distinct polar character is further reflected in the dielectric properties (Figure 3i). While BTH doping only mildly suppresses the maximum permittivity and introduce limited frequency dispersion (Figure S6a), it profoundly alters the response of the in-situ poled samples, which show a marked suppression of $\varepsilon_r$ (indicated by arrows in Figure 3i). This occurs despite a high relaxor diffuseness ($\gamma$ =1.97/1.98 for $x$ = 0.18/0.20, Figure S6b), highlighting the unique flexibility and field-responsiveness for the PNDs. Furthermore, the presence of thermally stimulated depolarization current (TSDC) peaks in all compositions signifies sustained coherence of the polar order (Figure 3j), despite the slush-like nanodomain morphology. This contrasts with the isolated PNRs typical of conventional relaxors and underscores the advantage of using high-polarizability cation for frustrating long-range order into PNDs without destruction of local polarity. In line with relaxor model proposed by Qi et al., the retained ferroelectricity in, for example, $x$ = 0.20 arises from the synergistic interplay between cation displacements and the frustrated OOT framework.[16b] Consequently, the local-scale coupling between polar configurations and OOT emerges as the critical mechanism underpinning the robust EC performance.

## Atomic-scale polarization configuration for high polarity

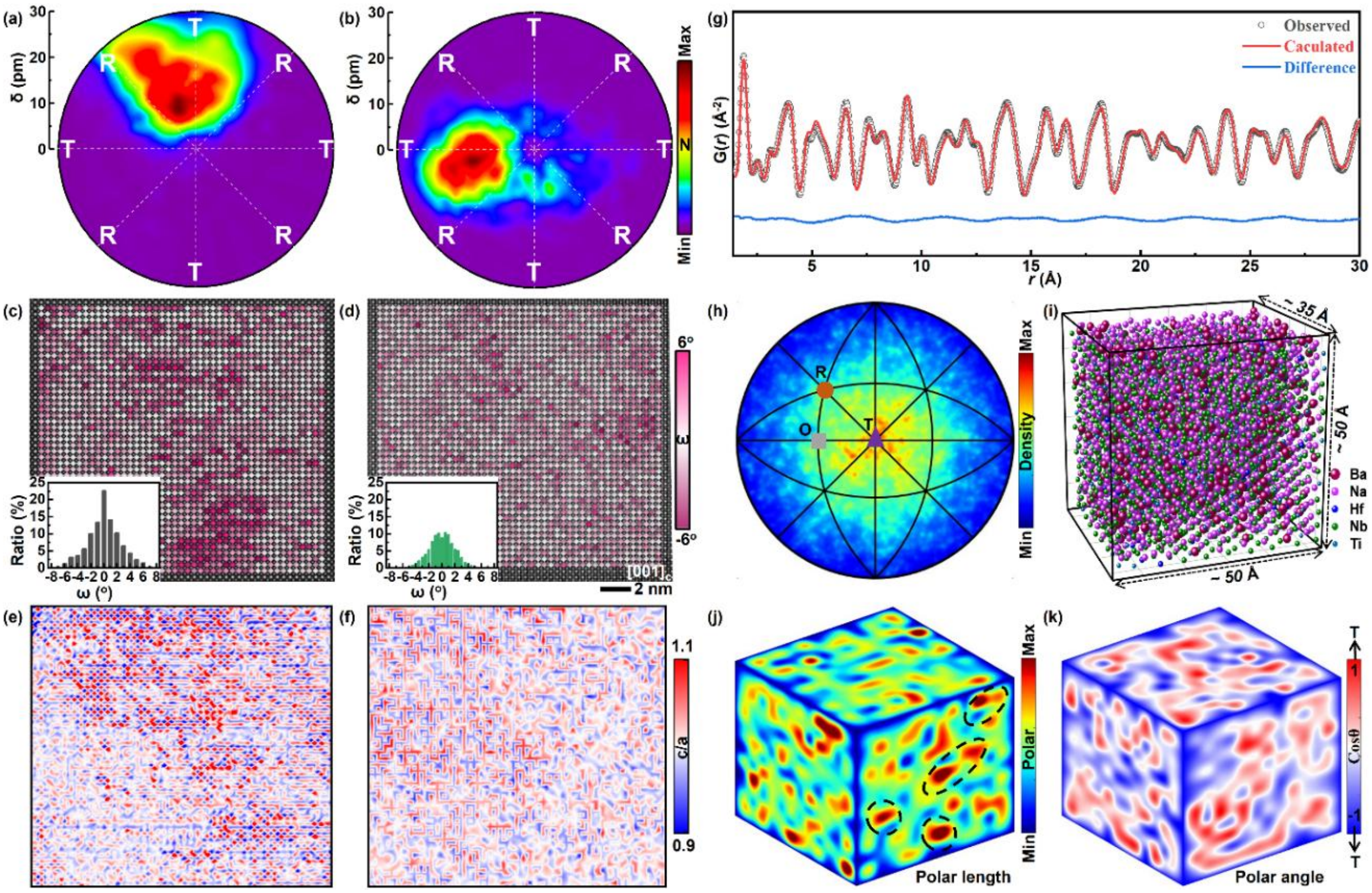

**Figure 4** a, b) Statistics of polarization vectors extracted from ABF-STEM along $[001]_c$ direction for $x$ = 0.14 and 0.20. Corresponding polarization vectors are displayed in Figure S7. The colors denote the numbers (N) for a certain direction. c, d) Atomic-resolution ABF-STEM image inlaid with OOT, and the anticlockwise and clockwise tilts of $BO_6$ octahedrons are marked for $x$ = 0.14 and 0.20, respectively. The statistical distributions of OOT angle are shown in their bottom-left insets. e, f) Maps of unit cell-level $c/a$ ratios for $x$ = 0.14 and 0.20. g) Fitting profiles of neutron pair distribution function (PDF) by performing reverse Monte Carlo (RMC) algorithm. h) Stereographic projections depicting the atomic polar displacement vectors for *A*/*B*-site cations. i-k) Distribution of the *A*/*B*-site atoms and corresponding unit-cell polarization vectors (including length and angle). $\theta$ denotes the angle of the polarization vector with respect to $[001]_c$.

To gain deeper insight into the local structure evolution, atomic-resolution annular bright-field scanning transmission electron microscope (ABF-STEM) is performed for $x$ = 0.14 and 0.20 compositions. Polar displacement vectors are overlaid using 2D Gaussian peak fitting based on the perovskite unit cell geometry (Figure S7). A clear contrast in polar configuration is observed: both compositions exhibit predominant polarization along the $[001]_c$ direction. This is consistent with macroscopic tetragonal symmetry, yet the magnitude of local polarization is notably suppressed in $x$ = 0.20 (Figure 4a, b). To resolve the OOT behaviour, ABF-STEM images are analyzed with overlaid schematic tilts showing periodic anticlockwise and clockwise rotations of the $BO_6$ octahedron (Figure 4c, d). The tilt angel $\omega$ is quantified as (180°−∠B-O1-B)/2 (see Figure 1a$_2$).[16c, 23] In $x$ = 0.14, the OOT pattern exhibit long-range order, whereas in $x$ = 0.20, the tilt become discrete and heterogeneous, accompanied by a broader distribution of $\omega$ (bottom-left insets, Figure 4c, d). This confirms that the polar wrench not only enhances local stress but also fragments both polar order and OOT correlation. The tetragonality (*c/a*), mapped from quantitative polarization displacement (Figure 4e, f), further highlights this transition. Unlike conventional relaxors where weakly polar regions are embedded in a pseudocubic matrix ($c/a \approx 1$),[13a, 24] $x$ = 0.14 shows strip-like regions of high *c/a,* which break into isolated yet strongly tetragonal patches in $x$ = 0.20. This indicates that the high-polarizability $Ba^{2+}$ and $Ti^{4+}$ cations disrupt the collective dipole alignment–forming PNDs without destroying the local polarization correlation. This is consistent with the dielectric and TSDC results.

To further resolve the local cation-driven polar order, neutron total scattering is performed on $x$ = 0.20. After extracting unit cell parameters from synchrotron XRD (SXRD) Rietveld refinement, a 9 × 9 × 9 supercell (≈ 50 Å × 50 Å × 35 Å) is constructed for reverse Monte Carlo (RMC) refinement, enabling reconstruction of the 3D atomic-scale nanostructure from total scattering data.[25] As displayed in Figure 4g, the real-space data G($r$) is well-fitted with high compatibility between experimental and fitting profiles. Based on the RMC analysis, the distribution of polar displacement directions along $[001]_c$ as stereographic projections is presented (Figure 4h). This is an effective means to differentiate the polar displacement vectors with different crystallographic symmetries. Apparently, the host cations exhibit a preference along the $[001]_c$ direction with a higher distribution density. The one-to-one 3D correspondence between the cation arrangement (Figure 4i) and unit-cell polarization is displayed (Figure 4j, k). Clearly, short-range coupling for polarization vectors is observed in $x$ = 0.20 strewn with highly-polar clusters (indicated by dash circles) and tetragonal phase populates the supercells. The revealed polar structure further validates the correlated polar clusters with *P4bm* phase.

## Structural attribute encoding temperature stability

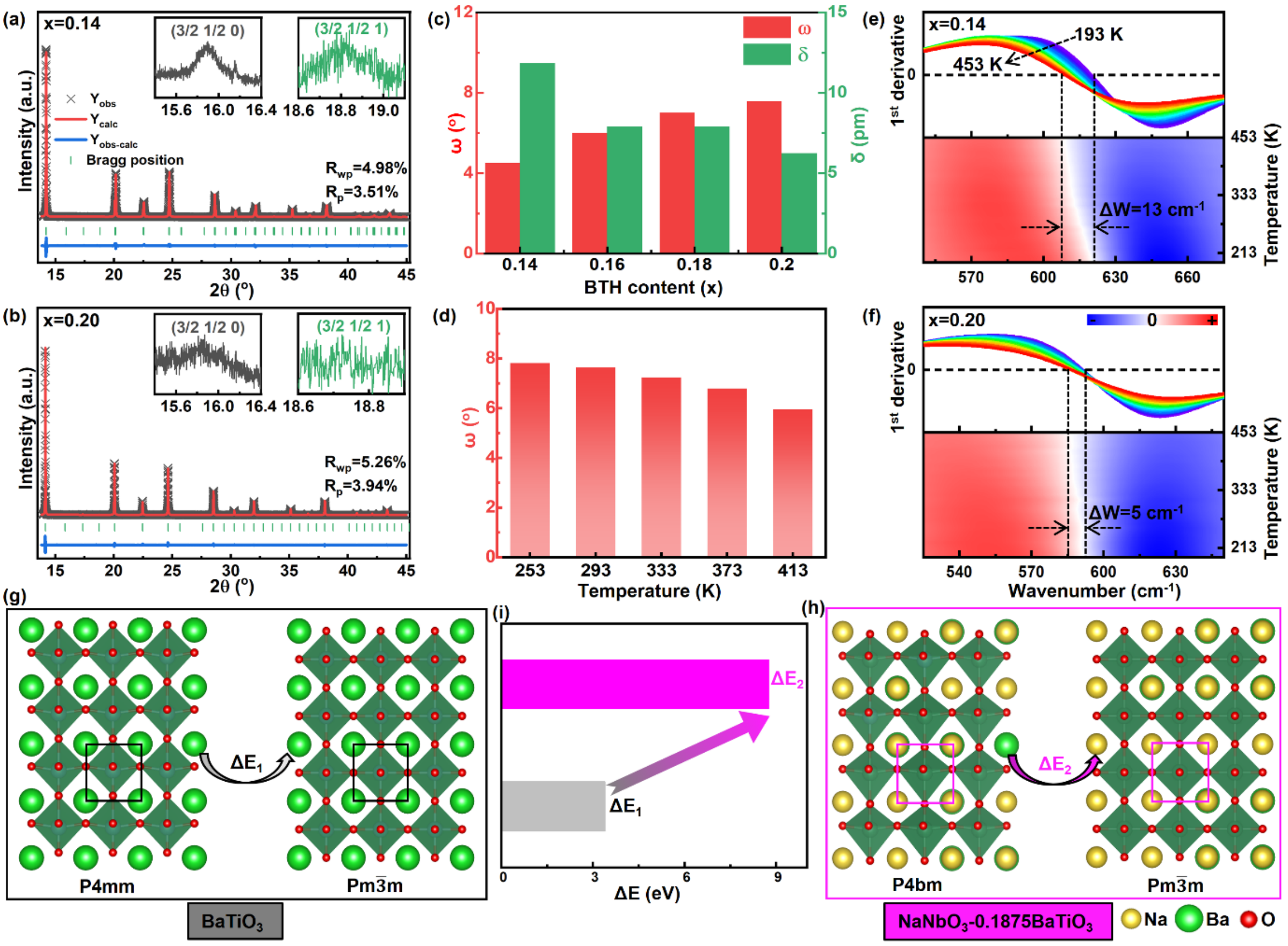

**Figure 5.** a, b) Experimental and Rietveld-refinement SXRD patterns of $x = 0.14$ and 0.20 ceramics; SXRD patterns for $x = 0.16$ and 0.18 are displayed in Figure S8 (Supporting Information), (3/2 1/2 0) and (3/2 1/2 1) superlattice reflections are shown in their insets. c) Evolution of OOT angle $\omega$ and *B*-site displacement $\delta$ as a function of BTH content at room temperature. d) Temperature dependent $\omega$ within 253 – 413 K for $x = 0.20$. e, f) The first (1st) derivative of Raman intensity to wavenumber of $BO_6$ octahedra vibration for $x = 0.14$ and 0.20 compositions, and $x = 0.16$ and 0.18 compositions are displayed in Figure S13 (Supporting Information). The 1st derivative value that equals to zero corresponds to peak position of $BO_6$ octahedra vibration. Full Raman spectra for $x = 0.16 – 0.20$ are displayed in Figure S14 (Supporting Information). g) The *P4mm* and *Pm*$\bar{3}$*m* phase models for $BaTiO_3$ (BT) and h) *P4bm* and *Pm*$\bar{3}$*m* phase models for $NaNbO_3$-0.1875$BaTiO_3$ (NN-0.1875BT). i) The energy barrier for *P4mm*-to-*Pm*$\bar{3}$*m* and *P4bm*-to-*Pm*$\bar{3}$*m* is denoted as $\Delta E_1$ and $\Delta E_2$. Obviously, $\Delta E_2$ is significantly higher than $\Delta E_1$. Considering the computational time-consuming and experimental compositions, the model of NN-0.1875BT is chosen (*P4bm* phase). Notably, the composition of NN-0.1875BT was very close to that designed $x = 0.20$ composition in this work.

The frustration of polar-order stems primarily from the enhanced stress field and exacerbated OOT induced by polar wrench effect. To further elucidate this structural response, SXRD are conducted across the $x = 0.14 – 0.20$ series (Figure 5a, b and Figure S8). All compositions exhibit the characteristic (3/2 1/2 0) and (3/2 1/2 1) superlattice reflections, unequivocally confirming the *P4bm* phase. Notably, the reduced intensity of these superlattice peaks in $x$ = 0.20 reflects the shortened OOT correlation length under increased wrenching effect. The average OOT angle $\omega$ and $B$-site displacement $\delta$, extracted from Rietveld-refinement, evolve systematically with $x$ (Figure 5c). The continuous increase in $\omega$ with $x$ supports the picture that polar frustration is driven jointly by OOT and local stress. This frustration generates abundant polar interfaces with high local polar entropy, compensating for the reduced enthalpy change typical for relaxors.[7a] Concurrently, the decreased $\delta$ lowers the energy barriers for polarization flipping, facilitating electric-field-induced alignment of polar nanodomains and thereby enhancing the EC response.

The transition from long-range order to polar nanodomains via polar frustration is directly reflected in the ferroelectric characteristics (Figure S9). With increasing $x$, $P$-$E$ loops evolve from square to slim, while a maximum polarization ($P_m$) is maintained for $x$ = 0.18 and 0.20. A recovery of normal ferroelectricity at lower temperature (Figure S10) further confirms that the polar configuration in these compositions can be reversibly regulated without degrading the intrinsic ferroelectricity–unlike the frozen behaviour often observed in $Bi_{0.5}Na_{0.5}TiO_3$-based relaxors at low temperature.[11] Most notably, the short-range, highly polar *P4bm* clusters in the $x$ = 0.20 exhibit exceptional thermal stability. No structural phase transition occurs up to 413 K, and both $\omega$ value (Figure 5d and Figure S11) and the characteristic superlattice reflections (Figure S12) remain largely unchanged. Raman spectroscopy provides further evidence: the first derivative of $BO_6$ vibration mode shows a minimal wavenumber shift ($\Delta$W = 5 $cm^{-1}$) for $x$ = 0.20 compared to other compositions (Figure 5g, h and Figure S13, S14), indicating stronger Nb-O bonding and local structural rigidity.[26] This resilience is a distinctive attribute for NN-based ferrodistortive relaxors and validates the initial design concept of using polar frustration for stable EC performance.

To shed light on the high temperature stability for $x$ = 0.20 sample, density functional theory (DFT) calculations are performed. A 4 × 4 × 3 supercell is constructed for *P4mm* and *P4bm* to investigate the energy barrier for phase transition, i.e., NN-0.1875BT with *P4bm*-to-$Pm\bar{3}m$, and pure BT with *P4mm*-to-$Pm\bar{3}m$ is selected for comparison (Figure 5g, h). As marked by square, the key difference between *P4bm* and *P4mm* phase is the singularity of OOT for the former. Here, the phase stability can be expressed by the energy difference ($\Delta E$) for phase transition between the ferroelectric (*P4mm*/*P4bm*) and paraelectric ($Pm\bar{3}m$) phase. As plotted in Figure 5i, the energy barrier $\Delta E_2$ for *P4bm*-to-$Pm\bar{3}m$ transition is ≈ 8.77 eV, which is 159% higher than that of *P4mm*-to-$Pm\bar{3}m$ ($\Delta E_1$ ≈ 3.39 eV). This result suggests phase transition in NN-0.1875BT is more difficult than conventional BT ceramics (without OOT), and thus underscores the critical role of *P4bm* network with OOT in resisting thermal degradation. Therefore, highly-polar frustrated nanodomains in ferrodistortive relaxor designed in this work provides a potent roadmap toward achieving high and thermally robust EC performances.

## 3. Conclusion

In summary, we introduce a feasible polar frustrations strategy, and highly-polar ferrodistortive nanodomains are established in $NaNbO_3$-Ba(Hf, Ti)$O_3$ relaxor. By utilizing this strategy, a large $\Delta T$ of 0.85 K and 0.70 K with an ultrawide $\Delta T_{span}$ of 118 K and 130 K is obtained for $x$ = 0.18 and 0.20 and thereof contributes to an ultrahigh figure of merit of > 90 $K^2$, significantly outperforms the lead-based/lead-free counterparts. Combined with 2D information from atomic-resolution STEM, 3D big-box model constructed from neutron total scattering and density functional theory calculations, we demonstrated that these highly-polar ferrodistortive nanodomains are responsible for the superior EC performances. The abundant highly-polar orders enhance entropy change while the robust OOT framework resists thermal perturbations. Since polar order fundamentally governs ferroelectric properties, our design framework provides a viable path for developing electrically active materials through construction of highly-polar frustrated nanodomains for applications in entropy-change-based ferroelectric and ferromagnetic materials beyond.

## 4. Experimental Section

*Sample Preparation*: High-purity raw chemicals $Na_2CO_3$, $Nb_2O_5$, $BaCO_3$, $TiO_2$, and $HfO_2$ were used to fabricate (1-$x$)$NaNbO_3$-$x$Ba($Ti_{0.89}Hf_{0.11}$)$O_3$ ($x$ = 0.14, 0.16, 0.18, and 0.20) ceramics via a conventional solid-state method. Raw $Na_2CO_3$ powder was desiccated at 423 K to remove the absorbed moisture and 0.8wt%$MnCO_3$ was added to optimize the sintering behaviors. Mixed raw chemicals were ball-milled for 12 h and calcined twice at 1171 and 1213 K for 3 h. Sintering temperature ranged in 1533 – 1543 K. Polished pellets were pasted with silver electrodes for electrical characterizations.

*Structure Characterizations*: Temperature-dependent synchrotron XRD (SXRD) were collected at the Australian Synchrotron Beamline ($\lambda$ = 0.6893 Å). Bright-field images were performed by transmission electron microscopy (TEM, FEI TalosF200X, US). In situ Raman spectra were collected by a Raman scattering spectrometer (Horiba Jobin-Yvon HR800, France) with a 532 nm laser. Domain amplitude and phase images were obtained using piezoelectric force microscopy (PFM, Cypher ES, Asylum Research). Atomic-scale images were acquired using an atomic resolution STEM (FEI Titan Themis 300). Accurate atomic positions in the STEM images were determined by 2D Gaussian peak fitting. The polarization configurations and polarization angle were calculated by customized MATLAB scripts. Neutron powder diffraction patterns were collected using the Multi-Physics Instrument at China Spallation Neutron Source (CSNS) in Dongguan, China. The powder samples were measured at room temperature for 2 hours in the time-of-flight (T.O.F.) mode with wavelength bands of 1 ~ 30 Å. The total neutron scattering data were processed using the Rzera software and Fourier transformation to obtain the pair distribution functions with a maximum momentum transfer $Q_{max}$ = 50 $Å^{-1}$.[27]

*Electric Property Measurements*: Dielectric properties were monitored by TH2832 (Tonghui Co., China) for in situ poled and fresh samples within 1–100 kHz with heating rate of 3 K $min^{-1}$. Temperature dependent *P–E* loops were measured by a ferroelectric test system (Precision LC, Albuquerque, NM) with a temperature controller (Delta-9023, Delta design, US) during cooling process from 423–213 K. Thermally stimulated depolarization current profiles were collected by using 6517B source meter (Keithley, US). Directly-measured EC data was collected by a homemade device with a PT100 resistor. The COMSOL Multiphysics software is used to build an equivalent test model and to simulate thermal equilibrium process.[11]

*Theoretical Calculations*: Density functional theory (DFT) is performed to study the temperature stability for $x$ = 0.20 sample. The Vienna Ab initio Simulation Package (VASP) with projector augmented wave (PAW) method are used to perform the DFT calculation.[28] The exchange-correlation effects are treated with the Perdew-Burke-Ernzerhof (PBE) generalized gradient approximation (GGA).[29] The plane-wave cutoff energy is set to be 500 eV, and the Brillouin zone is sampled with a 1×1×3 Monkhorst-Pack k-point grid. Electronic self-consistent calculation is converged to within $10^{-6}$ eV, while ionic relaxation is proceeded until the forces on all atoms are below 0.01 eV/Å.[30] Considering the computational time-consuming and experimental compositions, the model of 0.8125$NaNbO_3$-0.1875$BaTiO_3$ (NN-0.1875BT) was chosen (*P4bm* phase). Notably, the composition of NN-0.1875BT was very close to that designed $x$ = 0.20 composition in this work.

**Supporting Information**

Supporting Information is available from the Wiley Online Library or from the author.

**Acknowledgements**

This work was supported by the National Natural Science Foundation of China (Grant Nos. 12104001 and 12174001). The authors are grateful for the support from the Australian Synchrotron and thank Dr. Wenliang Tan for his help during the experiments at the Australian Synchrotron.

**Conflict of Interest**

The authors declare no conflict of interest.

**Data Availability Statement**

The data that support the findings of this study are available from the corresponding author upon reasonable request.

# Supporting Information

## Robust Electrocaloric Performance Enabled by Highly-Polar Frustrated Nanodomains in $NaNbO_3$-Based Ferrodistortive Relaxor

Feng Li, Changshun Dai, He Qi*, Jiecheng Liu, Xiaoming Shi*, Heng Zhou, Qiong Yang*, Mingsheng Long, Lei Shan, Chunchang Wang, Jianli Wang, Zhenxiang Cheng*

F. Li, C.S. Dai, M.S. Long, L. Shan

Institutes of Physical Science and Information Technology, Leibniz International Joint Research Center of Materials Sciences of Anhui Province, Anhui University, Hefei 230601, China.

H. Qi

School of Materials Science and Engineering, Hainan University, Haikou, Hainan, 570228, China.

Email: qihe@hainanu.edu.cn

J.C. Liu, X.M. Shi

Department of Physics, University of Science and Technology Beijing, Beijing, 100083, China.

Email: xmshi@ustb.edu.cn

H. Zhou, Q. Yang

Hunan Provincial Key Laboratory of Thin Film Materials and Devices, School of Materials Science and Engineering, Xiangtan University, Xiangtan, Hunan 411105, China.

Email: qyang@xtu.edu.cn

C.C. Wang

Laboratory of Dielectric Functional Materials, School of Materials Science and Engineering, Anhui University, Hefei, 230601, China.

J.L. Wang, Z.X. Cheng

Institute for Superconducting and Electronic Materials, Australian Institute for Innovative Materials, University of Wollongong, North Wollongong 2500, Australia.

Email: cheng@uow.edu.au

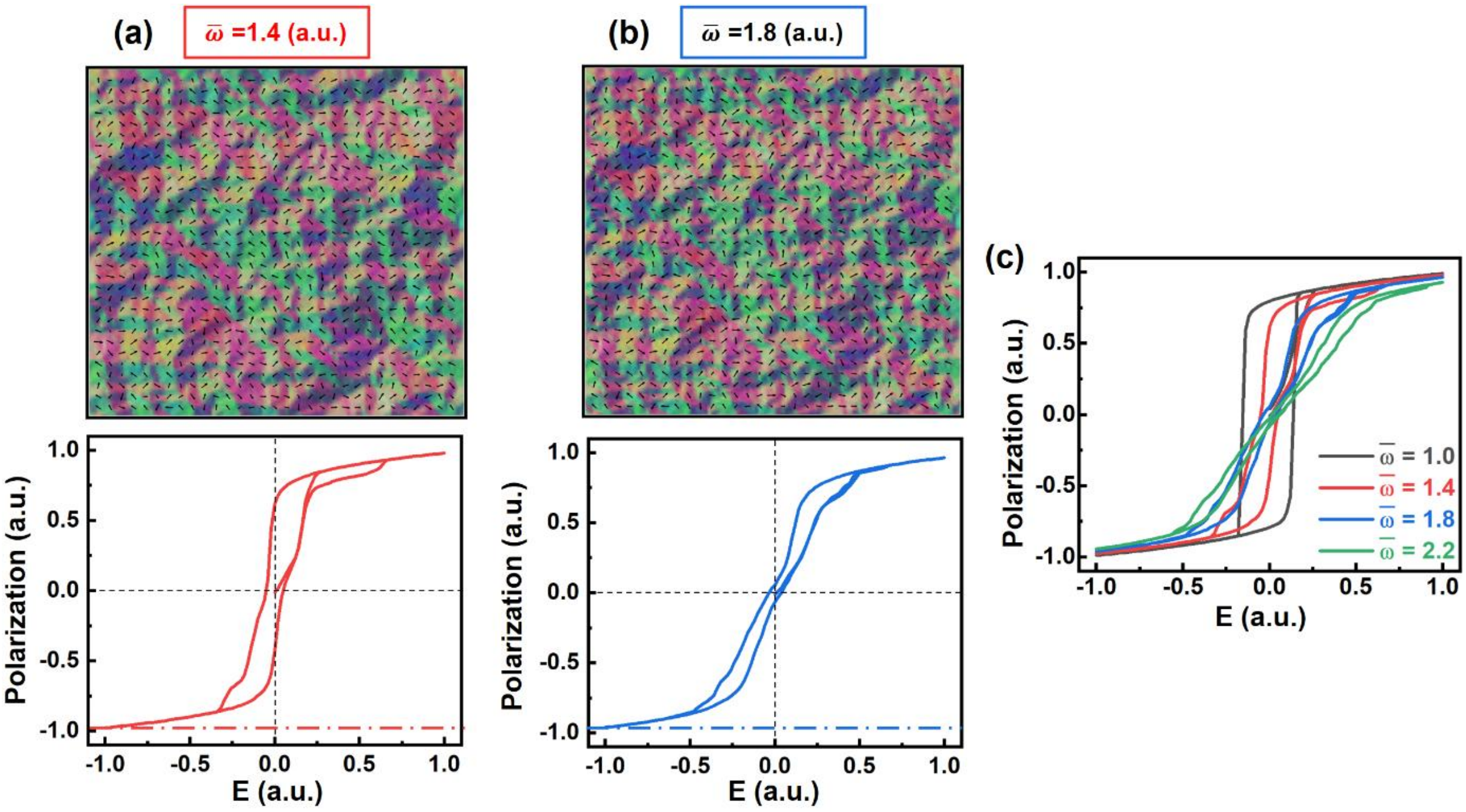

**Figure S1** a, b) Simulation of the 2D vector contours, phase distributions, and *P-E* loops for $\varpi$ = 1.4 and 1.8. c) Simulated *P-E* loops for $\varpi$ = 1.0 – 2.2 states.

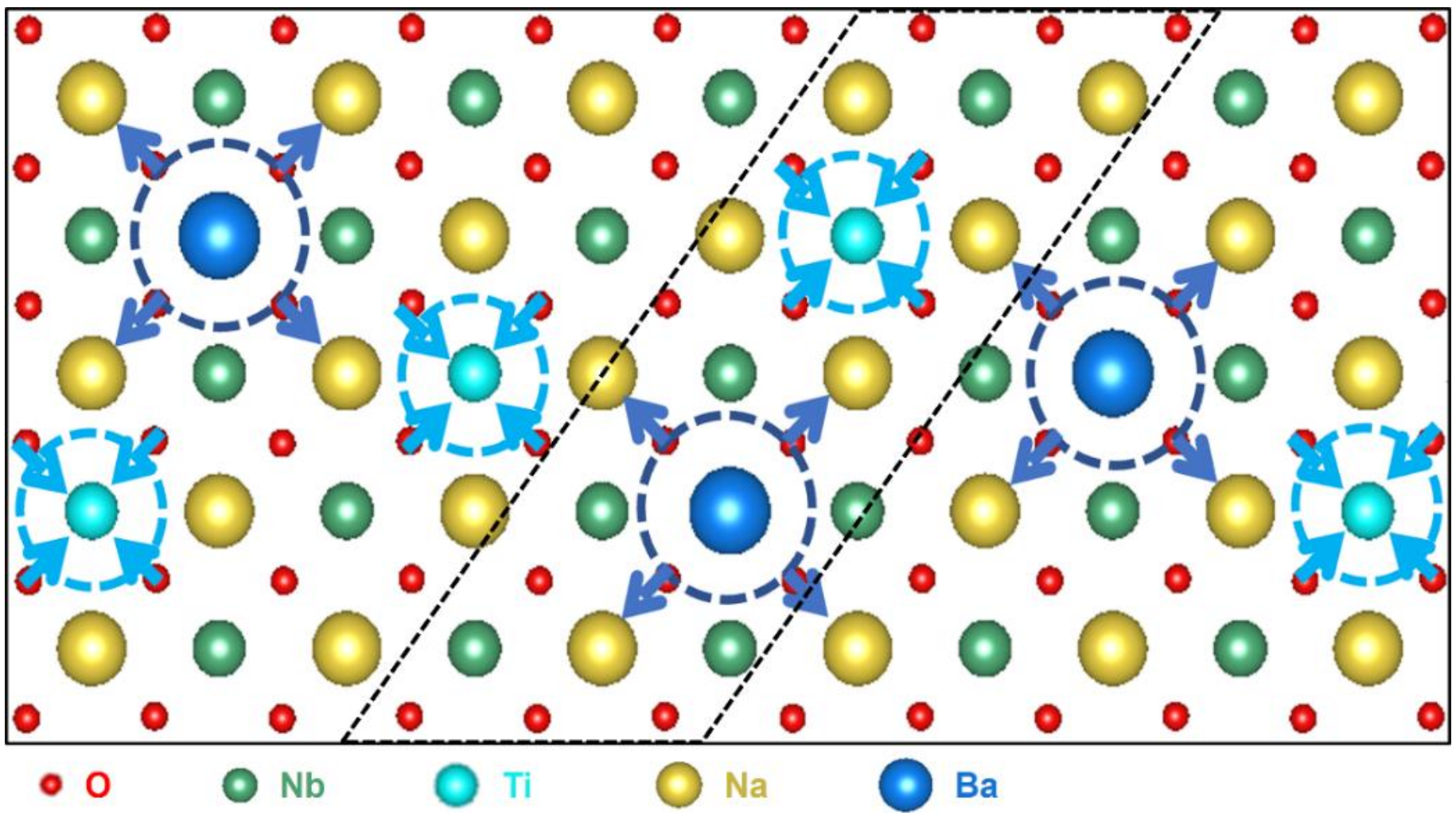

**Figure S2** Schematic graph for clarifying the roles of $Ba^{2+}$ and $Ti^{4+}/Hf^{4+}$ cation in engineering the lattice distortion. A larger size of $Ba^{2+}$ (1.61 Å) and smaller size of $Ti^{4+}$ (0.605 Å) is compared to $Na^{+}$ (1.39 Å) and $Nb^{5+}$ (0.64 Å), thus $Ba^{2+}$ imposes a compressive stress and $Ti^{4+}$ a tensile stress, respectively (due to the low concentration of Hf element, it is not displayed in this framework). This leads to a polar frustration.

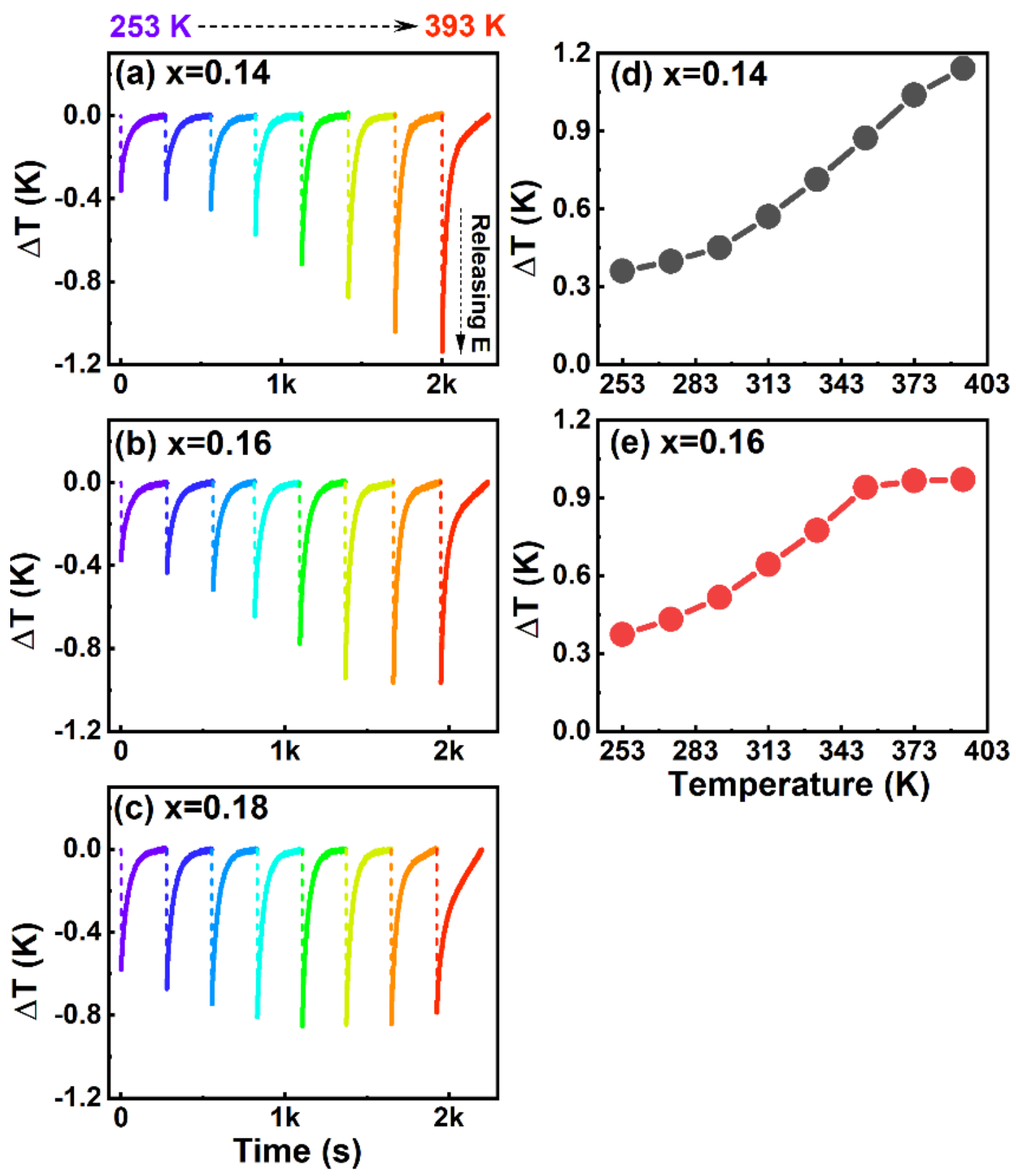

**Figure S3** a-c) Calibrated EC profiles for $x$ = 0.14–0.18. d, e) Temperature dependence of $\Delta T$ for $x$ = 0.14 and 0.16 samples.

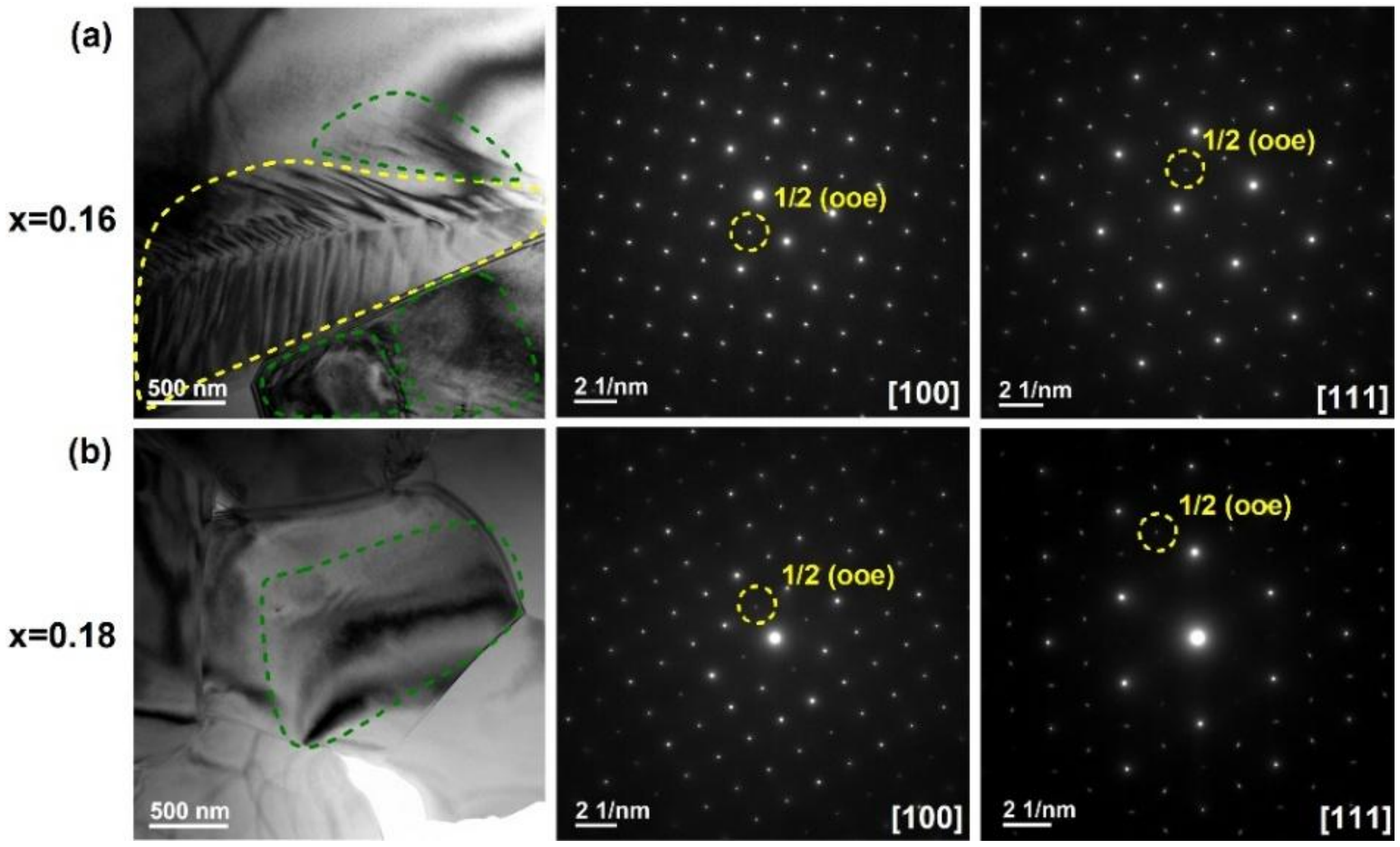

**Figure S4** a, b) TEM and SAED patterns for $x$ = 0.16 and 0.18 compositions.

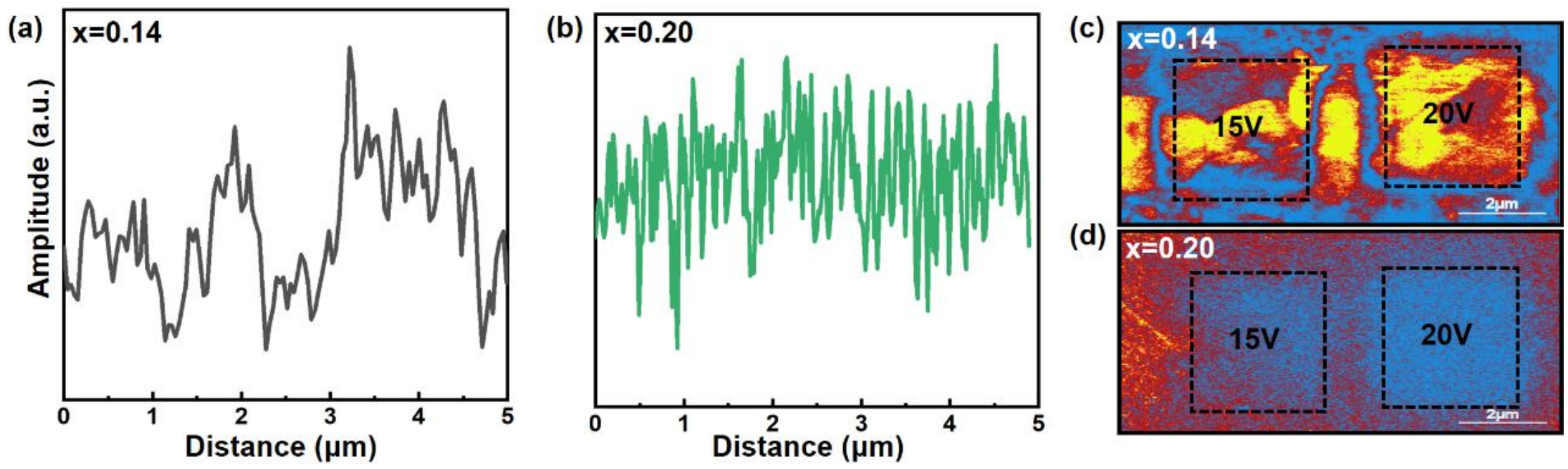

**Figure S5** a, b) Piezoresponse amplitude profiles for $x$ = 0.14 and 0.20. c, d) Local poling for $x$ = 0.14 and 0.20 with a voltage of 15 and 20 V within 3 × 3 μm$^2$ area.

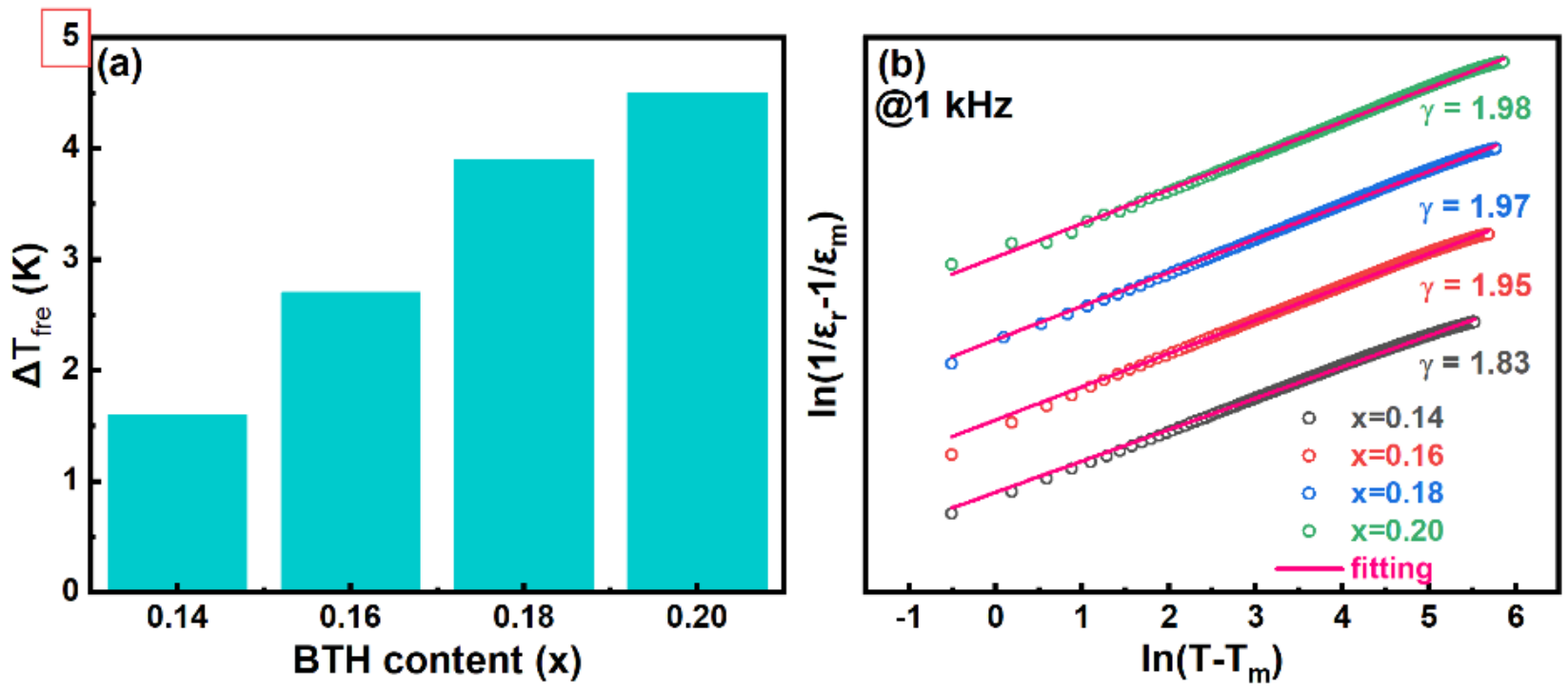

**Figure S6** a, b) Variation of $\Delta T_{fre}$ and (b) $ln(1/\varepsilon\text{-}1/\varepsilon_m)$ versus $ln(T\text{-}T_m)$ at 1 kHz for $x$ = 0.14 – 0.20 ceramics.

The relaxor characteristics of the ceramics are described by the degree of frequency dispersions $\Delta T_{fre}$ that defined as: $\Delta T_{fre}=T_m$ (100 kHz)-$T_m$ (1 kHz), where $T_m$(100 kHz) and $T_m$(1 kHz) denotes the value of $T_m$ at 100 kHz and 1 kHz, respectively.[S35]

The coefficient $\gamma$, defined as the degree of diffusiveness, which quantifies the deviation from the Curie-Weiss law: $1/\varepsilon - 1/\varepsilon_m = C^{-1}(T-T_m)^{\gamma}$, where γ ($1 \leq \gamma \leq 2$), $C$ is the Curie constant and $\varepsilon_m$ denotes the maximum dielectric constant.[S36]

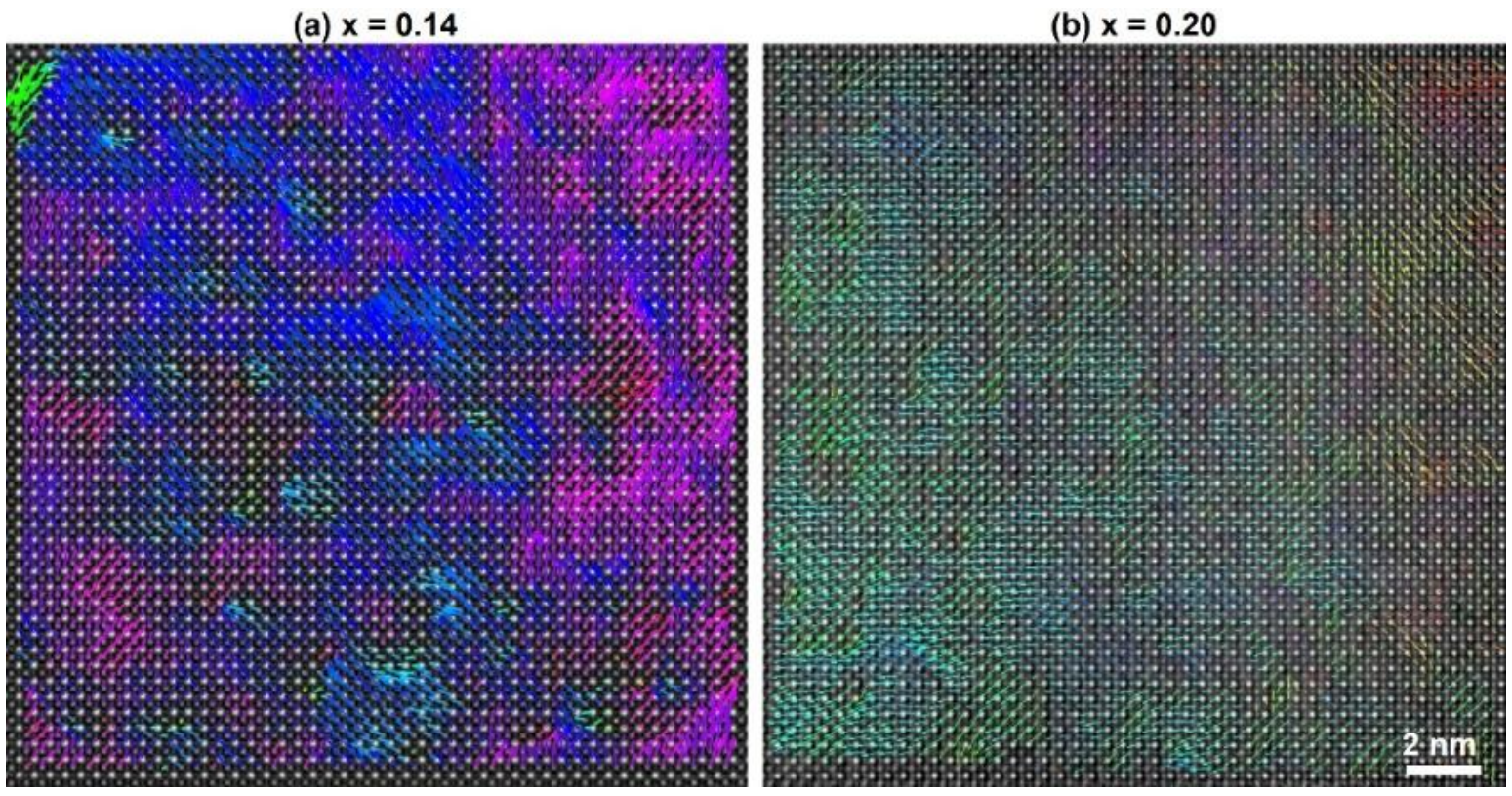

**Figure S7** a, b) Atomic-scale ABF-STEM polarization vectors mapping for $x$ = 0.14 and 0.20.

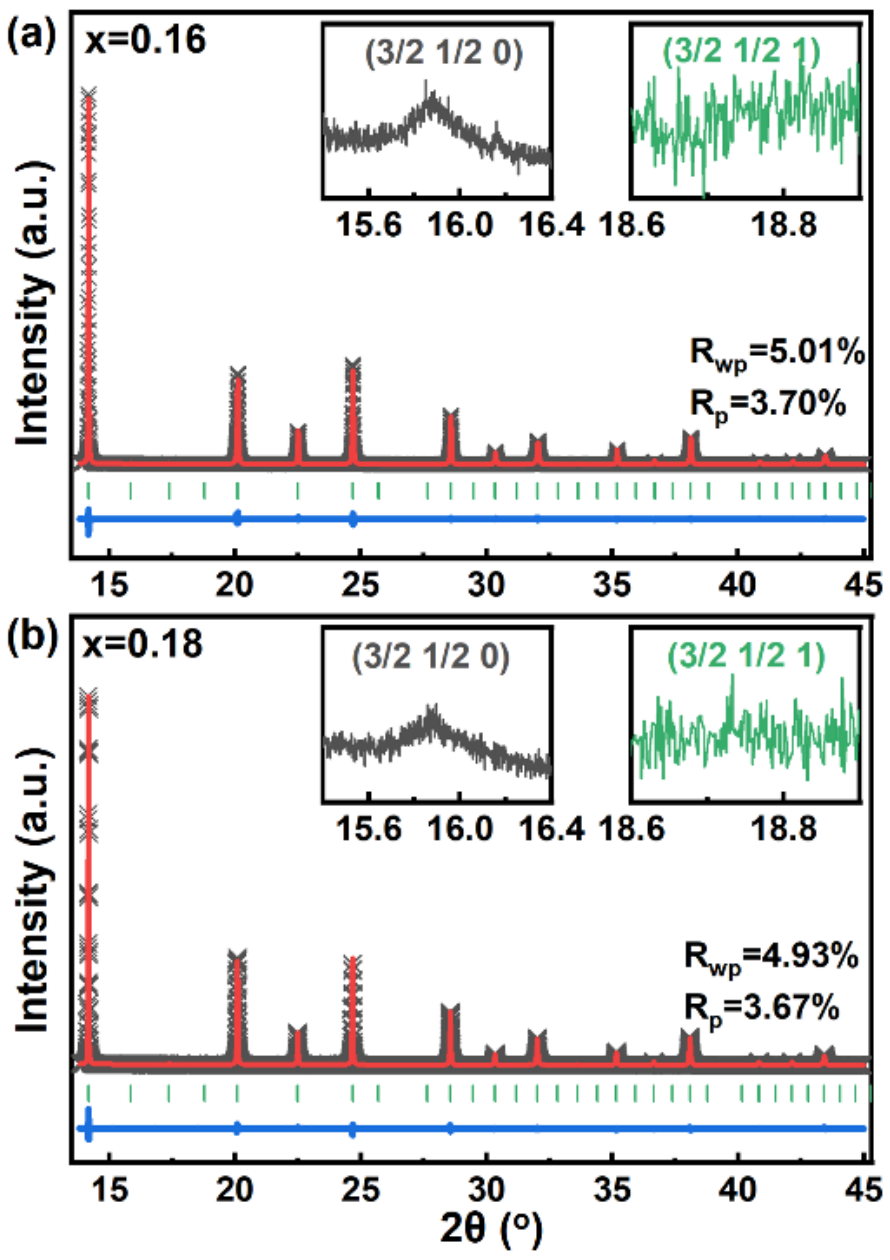

**Figure S8** a, b) SXRD patterns for $x = 0.16$ and 0.18, (3/2 1/2 0) and (3/2 1/2 1) superlattice reflections are shown in their insets.

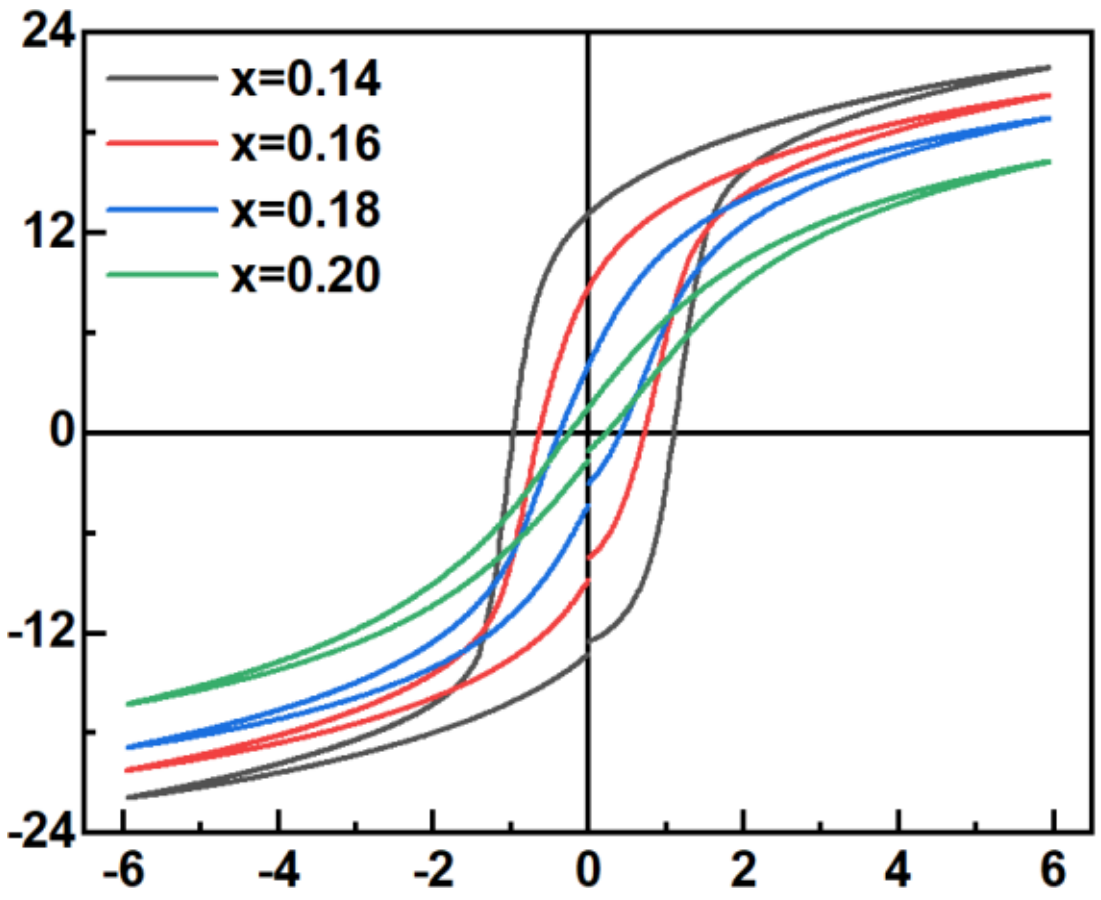

**Figure S9** Room-temperature *P*-*E* loops for $x = 0.14 – 0.20$.

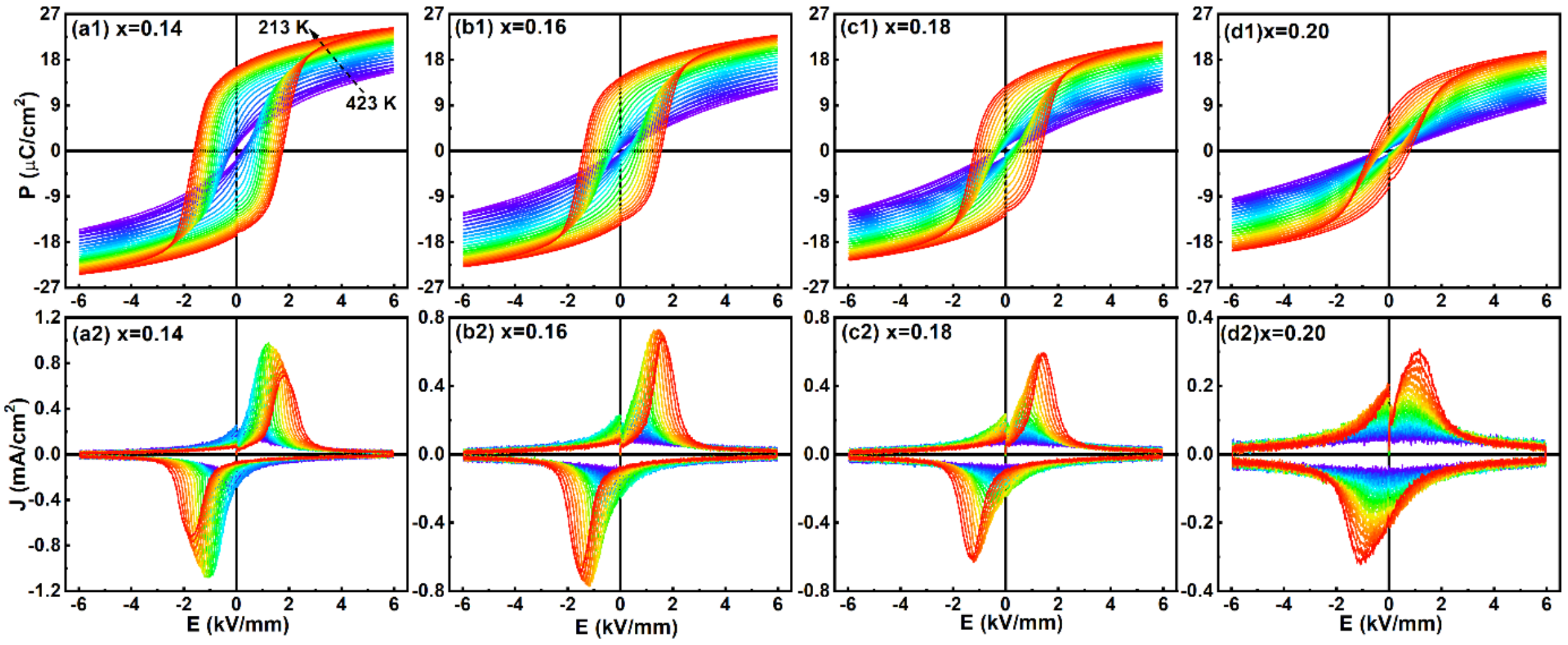

**Figure S10** Temperature dependent $a_1$-$d_1$) *P-E* and $a_2$-$d_2$) *J-E* loops for $x$ = 0.18 sample within temperature range of 213 – 423 K.

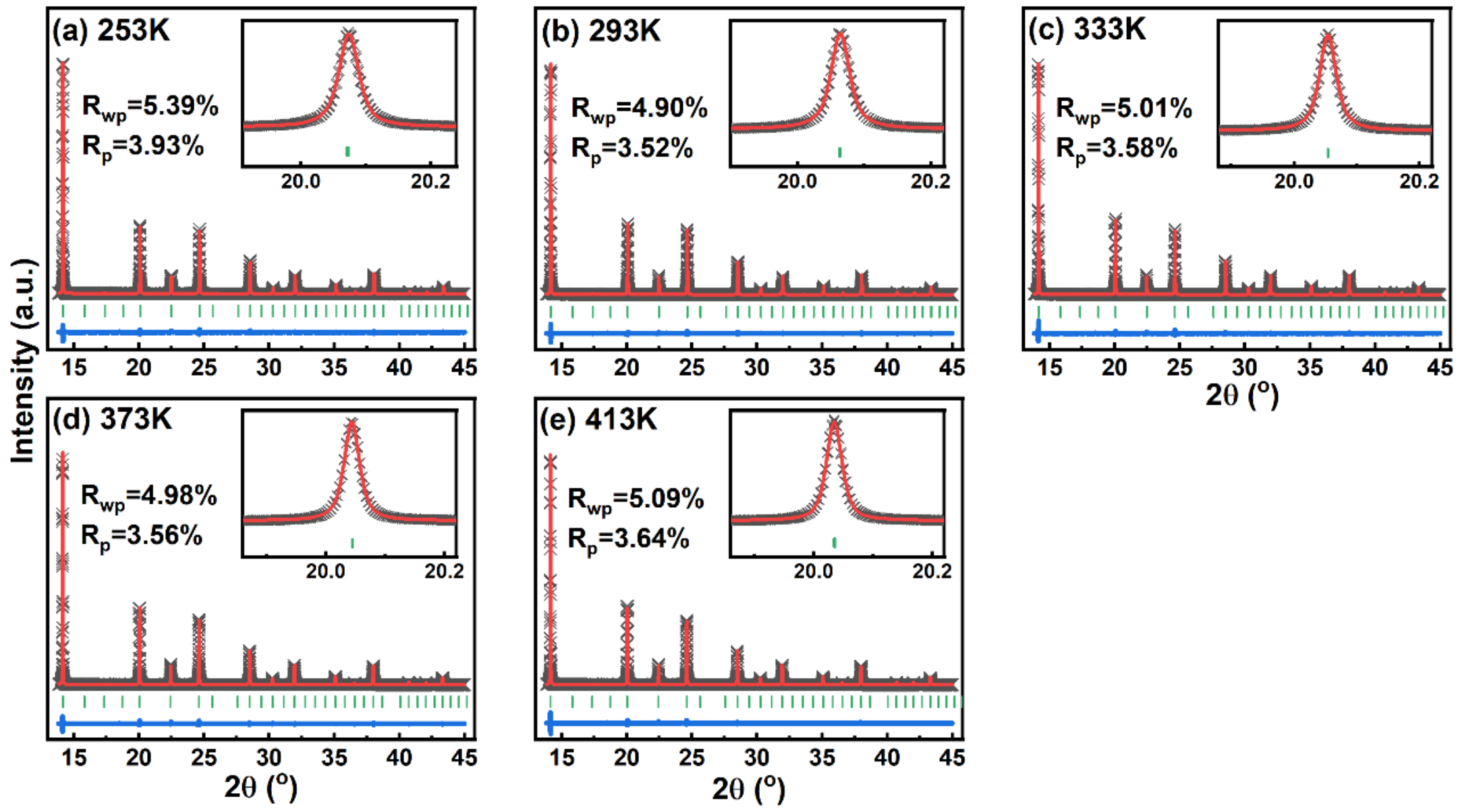

**Figure S11** (a-e) Experimental and Rietveld-refinement SXRD patterns of $x$ = 0.20 composition within the temperature range of 253 – 413 K.

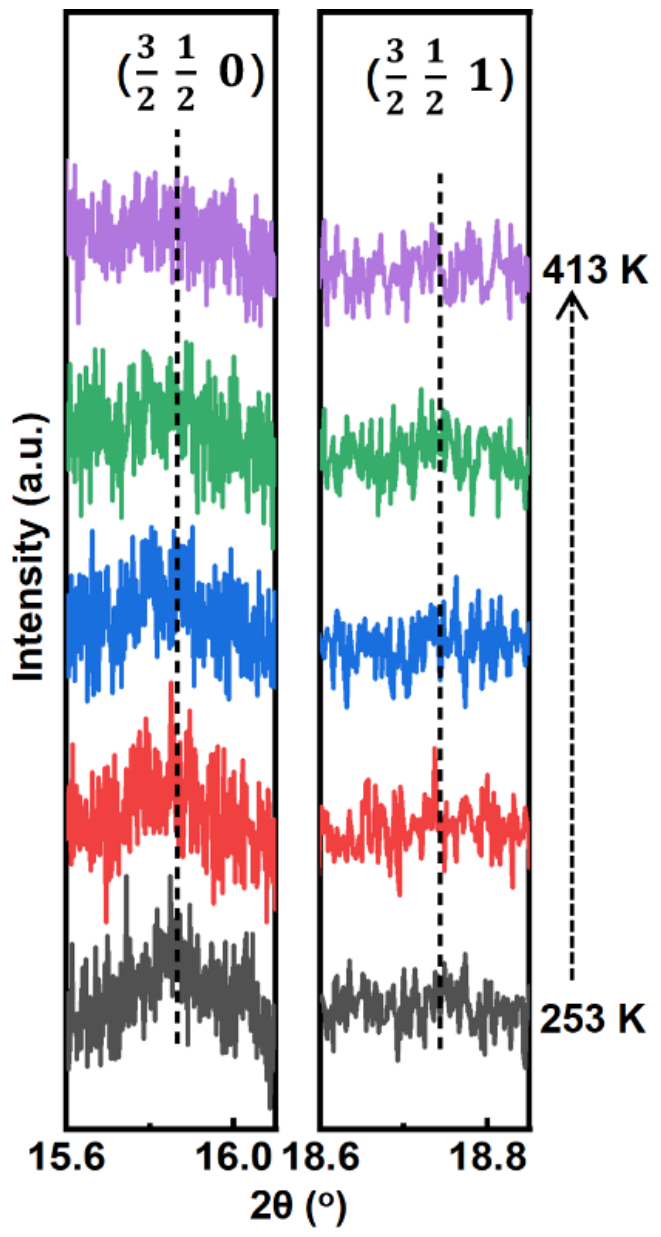

**Figure S12** (3/2 1/2 0) and (3/2 1/2 1) superlattice diffractions evolution for $x$ = 0.20 sample within 253 – 413 K.

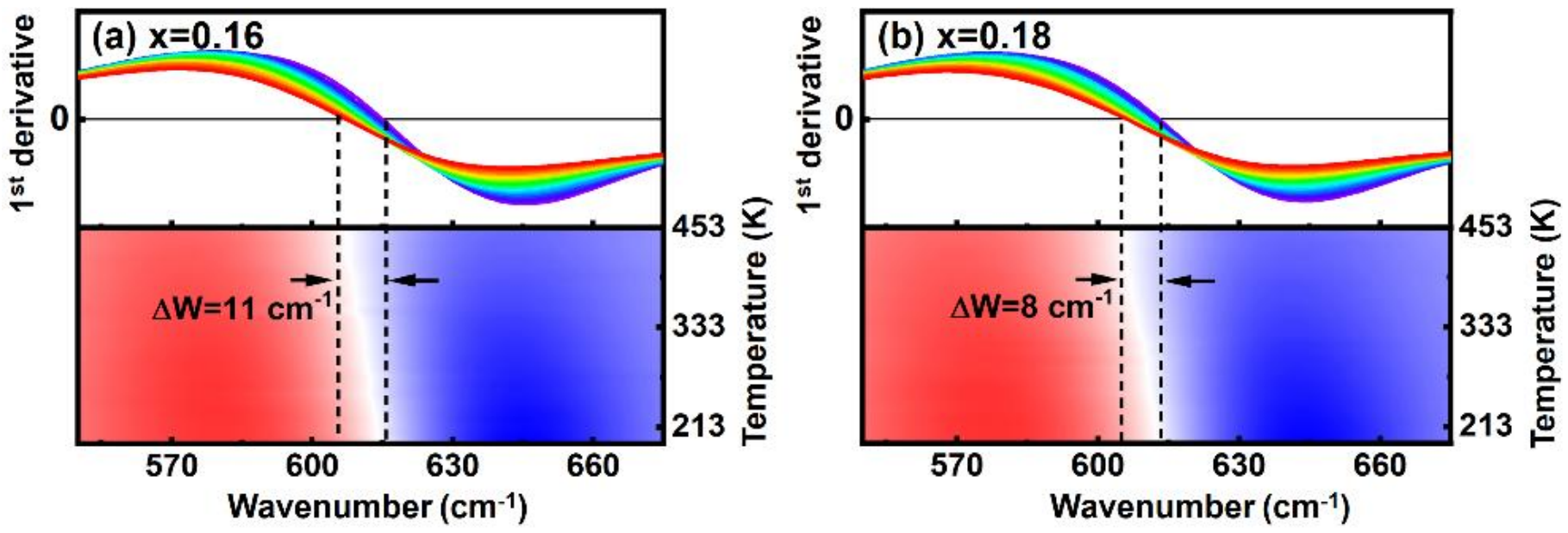

**Figure S13** The first (1st) derivative of Raman intensity to wavenumber of $BO_6$ octahedra vibration for $x$ = 0.16 and 0.18, the 1st derivative value that equals to zero corresponds to peak position of $BO_6$ octahedra vibration.

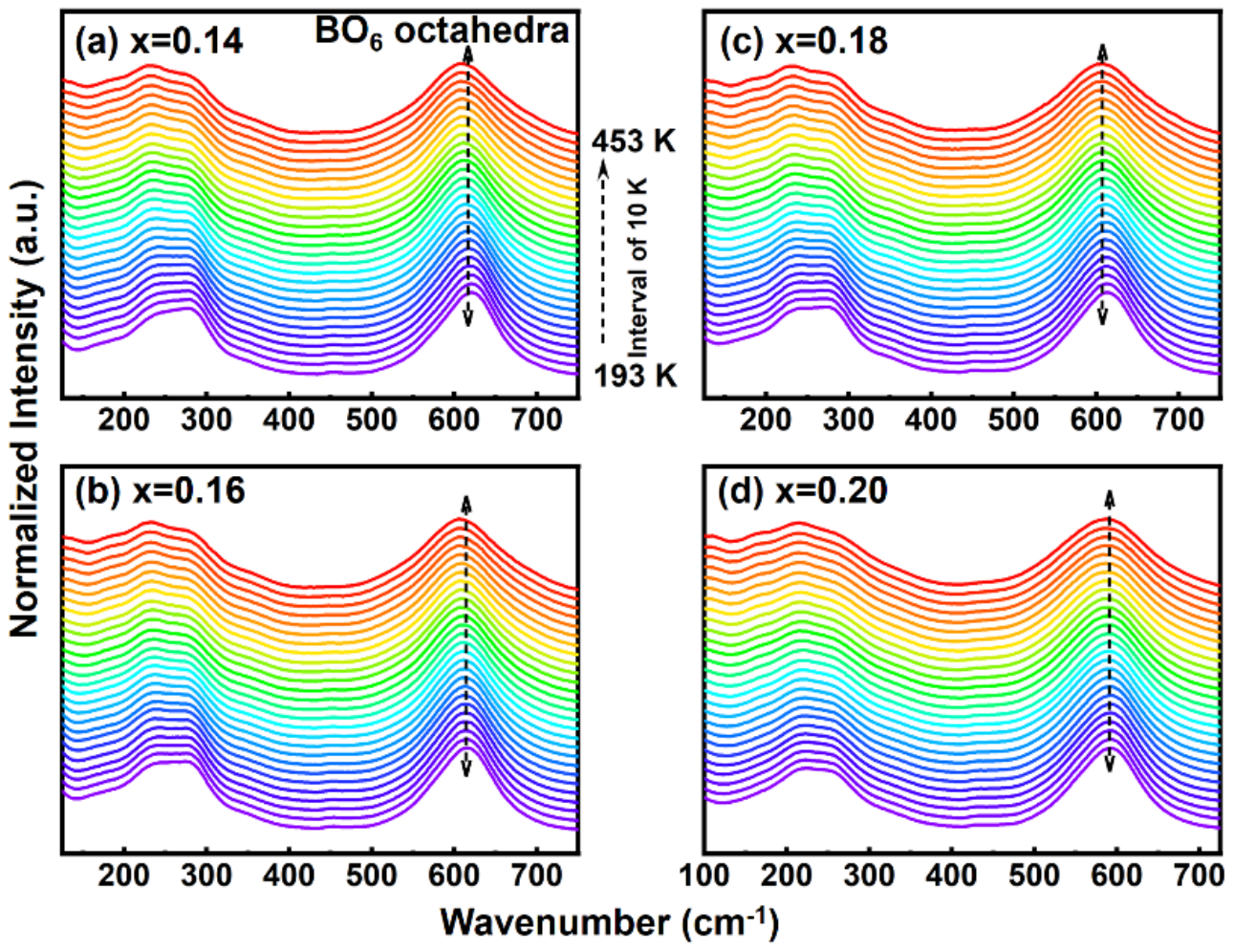

**Figure S14** (a–d) In situ Raman spectra for $x$ = 0.14–0.20 compositions.

**Note S1**

*Phase field simulation*: In this phase-field simulation, the polarization field $\mathbf{P} = (p_1, p_2, p_3)$ is selected as order parameter. The polarization field's temporal evolution is solved by the time-dependent Ginzburg-Landau (TDGL) equation:

$$\gamma_{ij} \frac{\partial P_i}{\partial t} + \frac{\delta F}{\delta P_j} = 0 \qquad \text{(Eq. S1)}$$

where $\gamma_{ij}$ is a kinetic coefficient that is related to the domain-wall mobility. $F$ is the total free energy of the system. $\frac{\delta F}{\delta P_j}$ is the thermodynamic driving force for polarization evolution, and $t$ is the time.[S37–S39]

The total free energy $F$ of the system is the sum of Landau energy ($F_{\text{Land}}$), elastic energy ($F_{\text{elas}}$), electric energy ($F_{\text{elec}}$), and gradient energy ($F_{\text{grad}}$). $f_{\text{Land}}$, $f_{\text{elas}}$, $f_{\text{elec}}$ and $f_{\text{grad}}$ are corresponding energy densities.

$$F = F_{\text{Land}} + F_{\text{elas}} + F_{\text{elec}} + F_{\text{grad}} = \iiint_V (f_{\text{Land}} + f_{\text{elas}} + f_{\text{elec}} + f_{\text{grad}})\text{d}V \qquad \text{(Eq. S2)}$$

The Landau free energy density:

$$f_{Land} = \alpha_1(P_1^2 + P_2^2 + P_3^2) + \alpha_{11}(P_1^4 + P_2^4 + P_3^4) + \alpha_{12}(P_1^2P_2^2 + P_2^2P_3^2 + P_1^2P_3^2) + \alpha_{111}(P_1^6 + P_2^6 + P_3^6) + \alpha_{112}[P_1^2(P_2^4 + P_3^4) + P_2^2(P_1^4 + P_3^4) + P_3^2(P_1^4 + P_2^4)] + \alpha_{123}P_1^2P_2^2P_3^2 + \alpha_{1111}(P_1^8 + P_2^8 + P_3^8) + \alpha_{1112}[P_1^6(P_2^2 + P_3^2) + P_2^6(P_1^2 + P_3^2) + P_3^6(P_1^2 + P_2^2)] + \alpha_{1122}(P_1^4P_2^4 + P_2^4P_3^4 + P_1^4P_3^4) + \alpha_{1123}(P_1^4P_2^2P_3^2 + P_2^4P_3^2P_1^2 + P_3^4P_1^2P_2^2)$$

(Eq. S3)

where $\alpha_1$, $\alpha_{11}$, $\alpha_{12}$, $\alpha_{111}$, $\alpha_{112}$, $\alpha_{123}$, $\alpha_{1111}$, $\alpha_{1112}$, $\alpha_{1122}$, and $\alpha_{1123}$ are the Landau coefficients. The electrostatic energy density $f_{elec} = -E_iP_i - \frac{1}{2}E_i^dP_i$, where $E_i$ is the applied electric field, and $E_i^d$ is the depolarizing electric field.

The elastic energy density:

$$f_{\text{elas}} = \frac{1}{2}c_{ijkl}e_{ij}e_{kl} = \frac{1}{2}c_{ijkl}(\varepsilon_{ij} - \varepsilon_{ij}^0)(\varepsilon_{kl} - \varepsilon_{kl}^0) \qquad \text{(Eq. S4)}$$

where $c_{ijkl}$ is the elastic stiffness tensor; $e_{ij}$, $\varepsilon_{ij}$ and $\varepsilon_{ij}^0$ represent the elastic strain, the total elastic strain, and the eigenstrain, respectively. The eigenstrain is calculated by $\varepsilon_{ij}^0 = Q_{ijkl}p_kp_l$, where $Q_{ijkl}$ is the electrostrictive coefficient.

The gradient energy density: $f_{\text{grad}} = \frac{1}{2}g_{11}\left(P_{1,1}^2 + P_{2,2}^2 + P_{3,3}^2\right) + g_{12}\left(P_{1,1}P_{2,2} + P_{2,2}P_{3,3} + P_{1,1}P_{3,3}\right) + \frac{1}{2}g_{44}\left[\left(P_{1,2} + P_{2,1}\right)^2 + \left(P_{2,3} + P_{3,2}\right)^2 + \left(P_{1,3} + P_{3,1}\right)^2\right] + \frac{1}{2}g'_{44}\left[\left(P_{1,2} - P_{2,1}\right)^2 + \left(P_{2,3} - P_{3,2}\right)^2 + \left(P_{1,3} - P_{3,1}\right)^2\right]$ (Eq. S5) where $g_{11}, g_{12}, g_{44}, g'_{44}$ are the gradient energy coefficients and $P_{i,j} = \frac{\partial P_i}{\partial x_j}$, where $x_j$ is the spatial coordinate.[S40]

In our simulation, the Landau coefficients are selected as:[S42]

$\alpha_1 = 3.61947 \times 10^5$ (T−345) $C^{-2}$ $m^2$ N, $\alpha_{11} = 0.9 \times 10^8$ $C^{-4}$ $m^6$ N, $\alpha_{12} = 8.0 \times 10^8$ $C^{-4}$ $m^6$ N, $\alpha_{111} = 3.3 \times 10^9$ $C^{-6}$ $m^{10}$ N, $\alpha_{112} = -3.5 \times 10^9$ $C^{-6}$ $m^{10}$ N, $\alpha_{123} = -1.0 \times 10^9$ $C^{-6}$ $m^{10}$ N, $\alpha_{1111} = 3.1 \times 10^{10}$ $C^{-8}$ $m^{14}$ N, $\alpha_{1112} = 0.2 \times 10^{10}$ $C^{-8}$ $m^{14}$ N, $\alpha_{1122} = 4.2 \times 10^{10}$ $C^{-8}$ $m^{14}$ N, $\alpha_{1123} = -5.0 \times 10^{10}$ $C^{-8}$ $m^{14}$ N, where $T$ is the temperature.

Other parameters are: $c_{11} = 2.30 \times 10^{11}$ N $m^{-2}$, $c_{12} = 0.90 \times 10^{11}$ N $m^{-2}$, $c_{44} = 0.76 \times 10^{11}$ N $m^{-2}$, $Q_{11} = 0.16$ $C^{-2}$ $m^4$, $Q_{12} = -0.072$ $C^{-2}$ $m^4$, $Q_{44} = 0.084$ $C^{-2}$ $m^4$, $g_{11}$=3.2 × $10^{-11}$ $C^{-2}$ $m^4$N, $g_{44} = g'_{44}$ = 1.6 × $10^{-11}$ $C^{-2}$ $m^4$N. The grid size is 128*dx* × 128*dx*, where *dx* =1 nm.